\documentclass{article}
\usepackage{PRIMEarxiv}
\usepackage[utf8]{inputenc}
\usepackage[T1]{fontenc}
\usepackage{amsmath,amsfonts,nicefrac}
\usepackage{graphicx}
\graphicspath{{images/}}
\usepackage{float,subcaption}
\usepackage{booktabs,array,tabularx,makecell}
\usepackage{microtype}
\usepackage[numbers,sort&compress]{natbib}
\usepackage{url}
\usepackage{hyperref}
\usepackage{doi}
\hypersetup{
  pdftitle={Trans-Domain Digital Twin: Conceptual Foundations, Architecture, and Research Outlook},
  pdfauthor={Mansoorali Amiri},
  pdfsubject={Trans-Domain Digital Twin Architecture and Operational Coupling},
  pdfkeywords={Digital Twin, Trans-Domain Digital Twin, TDDT, TDOC, System-of-Systems, Operational Coupling},
  pdflang={en-US},hidelinks
}
\title{Trans-Domain Digital Twin: Conceptual Foundations, Architecture, and Research Outlook}
\author{Mansoorali Amiri\,\href{https://orcid.org/0009-0002-1894-8089}{\includegraphics[scale=0.06,alt={Open Researcher and Contributor ID (ORCID)}]{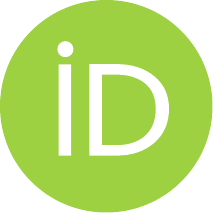}} \\
  Département d'informatique et de recherche opérationnelle (DIRO) \\
  Université de Montréal \\
  Montréal, QC, Canada \\
  \texttt{mansoorali.amiri@umontreal.ca}}

\begin{document}
\maketitle
\thispagestyle{empty}
\nocite{glaessgen2012digital,kritzinger2018digital,jones2020characterising,thelen2022comprehensivepart1,heindl2022structured,dalibor2022cross,verdouw2021digital,pylianidis2021introducing}
\nocite{escriba2024digital,vanbeek2023digital,bolender2021self,michael2022integration,wu2023comprehensive,mazzetto2024review,jiang2023multi,blair2021digital}
\nocite{david2025interoperability,fuller2020digital,gonzalez2025towards,amiri2025jumeaux,vergara2024federated,khedr2025composition,maatouk2017gaussian,kennedy2001bayesian}
\nocite{yang2019adaptive,modelica2024fmi302,ieee2025hla1516,wuni2026conceptualising,onaji2022digital,tao2018digital,trauer2020digital,rasheed2019digital_arxiv}
\nocite{rasheed2020digital,blanesvidal2008application,zuo2010fast,hutmacher2008computational,slotnick2023perspective,noy2001ontology,gruninger1995methodology,fernandezlopez1997methontology}
\nocite{uschold1995towards,euzenat2013ontology,matentzoglu2022simple,kantaros2025mathematical,raissi2019physics,orphanou2016dbnextended,okane2021dynamic,horrocks2004swrl}
\nocite{bonabeau2002agent,suetsugu2026microscopic,goldberg1989genetic,camacho2007model,shin2024applicability,thelen2022comprehensivepart2,mohammadi2024emulating,titsias2009variational}
\nocite{manousakas2020bayesian,polke2026adaptive,semenova2025case,birk2022automatic,mour2013agent,maier1998architecting,delaurentis2005understanding,ferko2022architecting}
\nocite{devrieze2024federated,shao2023credibility,sel2025survey,deantoni2024quantifying,taheri2022model,vanderlinden2019ligaps,tabase2023ammonia,nist2024digital}
\nocite{budiardjo2021digital,iso2021iso23247,correia2023data,karabulut2023ontologies,elhajj2024systematic,kustelega2024privacy,xia2024survey,yu2021model}
\nocite{xu2025survey,bray2024global,rockne2019mathematical,jarrett2018mathematical,hormuth2020forecasting,breuer2016colloquium,nathan2020universal,fowler2012surface}
\nocite{acharya2025quantum,litinski2019game,abdou2015blanket,taylor2023tritium,kasada2015system,bykerk2024retropropulsion,dresia2021multidisciplinary,guadagnini2022model}
\nocite{harirchi2022microbiological,batstone2002anaerobic,bernard2001dynamical,dudley2020competitive,kumar2022microbial,li2023model,hamelers2011butler,oladosu2016principle}
\nocite{holme2019induced,ma2021classical,gyagenda2022review,duberg2020ufomap,oleynikova2017voxblox,yao2023systematic,botinsanabria2022digital,ivanov2023intelligent}
\nocite{brucherseifer2021digital,burr2026realising,huang2022ethical}

\begin{abstract}
Complex systems comprise heterogeneous domains whose states, uncertainties, risks, and control consequences can cross domain boundaries. Existing cross-domain digital twin approaches broadly focus on comparison, reuse, semantic mapping, standardization, and interoperability, but do not inherently require operational connections among domain states, errors, objectives, constraints, decisions, and controls. This article proposes the trans-domain digital twin as an operational formulation along the continuum of Composite/Federated Digital Twin Systems. This approach connects heterogeneous domain twins through an aligned shared state, explicit coupling of data, models, states, errors, objectives, and controls, heterogeneous temporal coordination, joint decision-making, and feedback-based adaptation. The proposed framework presents a seven-layer conceptual architecture, a trans-domain orchestration core, minimum compliance conditions, a general operational formalism, progressive fast–meso–slow loops, and a single-episode offline training mechanism linked to bounded online adaptation. It also describes conceptual validation and evaluation criteria, a maturity model, a reference deployment architecture, and requirements for runtime safety, provenance, versioning, and model lifecycle management. The framework is conceptually mappable to standards for digital twins, model exchange, distributed simulation, and smart transducers; however, its formal compliance and operational effectiveness must be examined through independent benchmarks, uncertainty quantification, ablation testing, and field validation.

\end{abstract}


\section{Introduction}
\label{sec:introduction}
The initial objective of the Digital Twin (DT) was to provide a virtual representation of a physical system for monitoring, prediction, and decision-making throughout its life cycle \cite{glaessgen2012digital, kritzinger2018digital, jones2020characterising}. Despite the rapid growth of this concept across different industries, there is still disagreement regarding its definition, level of integration, role of real-time data, implementation method, and validation \cite{kritzinger2018digital, jones2020characterising, thelen2022comprehensivepart1}. Current studies show that the common patterns of these twins, such as simulation and optimization, are mainly focused on comparison and general development, and do not necessarily lead to operational coupling among heterogeneous domains \cite{heindl2022structured, dalibor2022cross}.

On the other hand, highly complex systems are inherently not single-domain, and a change in one domain alters the state and decisions of other domains \cite{verdouw2021digital, pylianidis2021introducing, escriba2024digital, vanbeek2023digital}. Accordingly, the Trans-Domain Digital Twin (TDDT) is defined in this article as a proposed formulation that connects multiple domain-specific twins through a shared state, coupled models, and decision-making loops \cite{verdouw2021digital, pylianidis2021introducing, vanbeek2023digital, bolender2021self}. The objective of this article is to clarify the distinction between TDDT and other twins, to present its conceptual architectural framework, and to outline a future pathway for the development of self-adaptive and verifiable trans-domain twins \cite{thelen2022comprehensivepart1, heindl2022structured, pylianidis2021introducing, bolender2021self}.

\subsection{The Evolution of Digital Twins from Single-Domain Systems to Multi-Domain Systems}

This evolution has been driven by the complexity of real-world systems and the need to integrate virtual entities and distributed twins \cite{jones2020characterising, michael2022integration, wu2023comprehensive}. In the single-domain approach, the initial focus was on monitoring, virtual representation, and data exchange between the physical system and the model within a limited domain, such as a machine or a building \cite{jones2020characterising}; however, due to the interdependence of the components of real-world systems, decisions cannot always be considered independently of other domains \cite{verdouw2021digital, escriba2024digital}. Although Cross-Domain studies have been conducted with the aim of identifying multi-faceted development patterns and enhancing interoperability across diverse domains, ranging from manufactured components to smart cities \cite{dalibor2022cross, mazzetto2024review}, the full realization of an integrated system requires a transition toward a “multi-domain and system-of-systems” approach, so that independent distributed twins can exchange shared information, objectives, and decisions by overcoming technological, data, model, and architectural heterogeneities \cite{michael2022integration, jiang2023multi, blair2021digital}.

This transition toward a trans-domain approach, in addition to the bidirectional flow of data, model, network, and application, requires operational connection of state, uncertainty, error, objective, decision, and control among domain-specific twins \cite{wu2023comprehensive}. Optimal, proactive, and real-time decision-making in today’s problems has a multi-objective and multi-domain nature and is better achieved only through a connected and trans-domain network, in order to prevent systemic errors arising from local and uncoordinated controls \cite{michael2022integration, wu2023comprehensive, jiang2023multi}.

\subsection{The Necessity of Introducing the Concept of the Trans-Domain Digital Twin}

In real-world multi-domain systems that are modeled using single-domain or separate digital twins, error, risk, and uncertainty in one domain can affect other domains; therefore, local decision-making and separate optimization of each domain may lead to an inconsistent or high-risk decision at the whole-system level \cite{michael2022integration, vanbeek2023digital, david2025interoperability}.

Cross-domain studies show that DTs in different domains have reusable common patterns, requirements, and mechanisms; in contrast, the literature on Digital Twin Systems-of-Systems raises the issue of composing and integrating multiple independent DTs at the level of larger systems.

However, the cross-domain approach mainly focuses on identifying common patterns, comparison, structured development, and reuse across domains \cite{dalibor2022cross, heindl2022structured}. Therefore, this article argues that a trans-domain conceptual design is necessary to cover the operational coupling of data, model, state, error, objective, and control, as well as to manage multi-scale systems, enable online adaptation with feedback from the real environment, and establish a clear framework for the classification, architecture, and validation of this generation of twins \cite{fuller2020digital, gonzalez2025towards, michael2022integration, vanbeek2023digital, david2025interoperability, amiri2025jumeaux}.

\subsection{Objective, Main Questions, and Structure of the Article}

The objective of this review is to clarify the position of TDDT as a proposed and emerging approach for the operational connection of domain-specific twins in complex systems. Accordingly, the article addresses four main questions:

\textbf{Q1. What is a trans-domain digital twin, and how does it differ from a single-domain, multi-domain, and Cross-Domain Digital Twin?}

This question is based on the existing ambiguity in the definition of DT, the difference in the level of data connection in the Digital Model, Digital Shadow, and Digital Twin, and the distinction between cross-domain reuse and trans-domain operational coupling in the proposed formulation of this article \cite{kritzinger2018digital, jones2020characterising, dalibor2022cross, heindl2022structured}.

\textbf{Q2. What is the position of TDDT in the classification, and why can it be considered along the continuum of Composite/Federated Digital Twin System-of-Systems?}

This question relies on studies that extend DT from the level of a single asset to the level of composable, multi-source, and system-of-systems structures, and examine the possibility of connecting multiple independent twins within a shared architecture \cite{dalibor2022cross, michael2022integration, vanbeek2023digital}.

\textbf{Q3. How should the conceptual architecture of TDDT connect data, model, state, error, objective, decision, and control among heterogeneous domains?}

This question arises from the need to integrate virtual entities, enable interoperability, manage heterogeneous data and models, and support decision-making based on a shared state in complex systems \cite{jones2020characterising, michael2022integration, david2025interoperability, wu2023comprehensive}.

\textbf{Q4. How can TDDT support trans-domain decision-making through fast inner loops and slow outer loops, offline learning, online tuning, and feedback from the real environment?}

This question is based on the logic of self-adaptive DTs, applications in smart agriculture, complex system design, and the trans-domain framework proposed in this article \cite{vanbeek2023digital, verdouw2021digital, amiri2025jumeaux}.

To answer these questions, the background and common classifications of digital twins are first reviewed; then, the concept of TDDT and its distinction from single-domain, multi-domain, and cross-domain approaches are explained. Next, the conceptual framework of TDDT, including domain-specific twins, the trans-domain shared state, types of coupling, fast and slow loops, and online adaptation mechanisms, is presented. Subsequently, selected applications of TDDT in different areas of complex systems are reviewed. Finally, the advantages, challenges, validation limitations, data and security issues, and future directions for the development of lightweight, self-adaptive, and trustworthy TDDTs are summarized.

\subsection{Contributions and Innovations of the Article}

The main contributions of this article are as follows: 1) defining TDDT as an operational formulation of a Composite/Federated Digital Twin System-of-Systems and distinguishing it from single-domain, multi-domain, and Cross-Domain twins \cite{heindl2022structured, dalibor2022cross, michael2022integration, david2025interoperability, vergara2024federated, khedr2025composition}; 2) presenting a seven-layer architecture and the Trans-Domain Orchestration Core (TDOC) for coordinating domain twins, the shared state, decisions, and feedback \cite{michael2022integration, amiri2025jumeaux, david2025interoperability, khedr2025composition}; 3) classifying trans-domain coupling into data, model, state, error, objective, and control, and defining fast, meso, and slow loops \cite{heindl2022structured, michael2022integration, amiri2025jumeaux, david2025interoperability, khedr2025composition}; 4) presenting a single-episode offline training and online adaptation mechanism for experience transfer, error correction, and reducing learning risk in the real environment \cite{bolender2021self, amiri2025jumeaux, maatouk2017gaussian, kennedy2001bayesian, yang2019adaptive}; and 5) explaining the conceptual mapping capability of the architecture with Functional Mock-up Interface (FMI) and High Level Architecture (HLA) \cite{modelica2024fmi302, ieee2025hla1516}.

The innovation of the article does not lie in the independent invention of all components, but rather in their integrated combination and formulation for operational trans-domain coupling, multiscale coordination, and joint decision-making \cite{heindl2022structured, dalibor2022cross, michael2022integration, amiri2025jumeaux, david2025interoperability, vergara2024federated, khedr2025composition}.
\subsection{Article Structure}

To answer these questions, the background and common classifications of digital twins are first reviewed; the concept of TDDT and its distinction from single-domain, multi-domain, and cross-domain approaches are then explained. Next, the conceptual framework of TDDT, including domain twins, the shared trans-domain state, coupling types, fast and slow loops, and online adaptation mechanisms, is presented. Selected applications of TDDT across different complex-system domains are then reviewed. Finally, the advantages, challenges, validation limitations, data and security issues, and future directions for developing lightweight, self-adaptive, and reliable TDDTs are summarized.

\subsection{Nomenclature and Abbreviations}

\begin{table}[ht]
\centering
\caption{Nomenclature and abbreviations}
\label{tab:abbreviations}
\small
\begin{tabularx}{\textwidth}{
>{\raggedright\arraybackslash}p{3.0cm}
>{\raggedright\arraybackslash}X
}
\textbf{Abbreviation} & \textbf{Full Form} \\[4pt]

DT & Digital Twin \\
CDDT & Cross-Domain Digital Twin \\
TDDT & Trans-Domain Digital Twin \\
SoS & System-of-Systems \\
TDOC & Trans-Domain Orchestration Core \\
$S_{\mathrm{TD}}$ & Shared Trans-Domain State \\
CRP & Context Reference Patterns \\
SARG & Stage-Aware Reference Guidance \\
MRG & Meso Reference-Guidance \\
SS-KStore & Step-Stream Memory Knowledge Store \\
SETD-KStore & Single-Episode Trans-Domain Knowledge Store \\
LCI & Loop Current Index \\
MPC & Model Predictive Control \\
FMI & Functional Mock-up Interface \\
FMU & Functional Mock-up Unit \\
HLA & High Level Architecture \\
ROM & Reduced-Order Model \\
PINN & Physics-Informed Neural Network \\
VVUQ & Verification, Validation, and Uncertainty Quantification \\
KStores & Knowledge Stores \\
UQ & Uncertainty Quantification \\
API & Application Programming Interface \\
AI & Artificial Intelligence \\
MAE & Mean Absolute Error \\
RMSE & Root Mean Square Error \\

\end{tabularx}
\end{table}

\section{Background and Current State of Digital Twins}
\label{sec:background}

The concept of the Digital Twin (DT) was initially introduced in connection with product lifecycle management and reducing the gap between a physical asset and its digital representation \cite{wuni2026conceptualising, glaessgen2012digital}. In the aerospace literature and later in smart manufacturing, this concept went beyond a simple information model and became a framework for combining operational data, health monitoring, simulation, prediction, and decision-making throughout the system lifecycle \cite{glaessgen2012digital, onaji2022digital, tao2018digital}. With the growth of IoT\footnote{Internet of Things}, CPS\footnote{Cyber–Physical Systems}, Industry 4.0, artificial intelligence, and data analytics, DT has become one of the main mechanisms for monitoring, prediction, and optimization in domains such as manufacturing, healthcare, smart cities, and smart agriculture \cite{tao2018digital, fuller2020digital, verdouw2021digital}. However, systematic reviews show that the definition, conceptual boundary, level of data connection, role of feedback, and method of DT validation remain diverse and sometimes inconsistent \cite{kritzinger2018digital, jones2020characterising, fuller2020digital}. On the other hand, recent studies show that the evolution of DT is moving from a single asset toward connected, multi-domain, interoperable, and system-of-systems (SoS) digital models; this trajectory provides the theoretical basis required for proposing trans-domain digital twins in this article \cite{dalibor2022cross, thelen2022comprehensivepart1}.

\subsection{General Definition of Digital Twin and Its Distinction from Digital Model and Digital Shadow}

A digital twin, in its general sense, is a digital representation of an object, machine, process, or physical system that uses data, models, computation, and, where available, feedback to reflect the behavior and state of the real system over time \cite{verdouw2021digital, trauer2020digital, rasheed2019digital_arxiv}. The main distinction among a Digital Model, a Digital Shadow, and a Digital Twin depends on the level of automation and the direction of data flow between the physical system and the digital model \cite{kritzinger2018digital}.

In a Digital Model, the digital representation is independent of the physical system, and data updates or effects on the real system are not performed automatically \cite{fuller2020digital, trauer2020digital, kritzinger2018digital}. In a Digital Shadow, data flow is established automatically but unidirectionally from the physical system to the digital model, and the model can reflect the real state but does not automatically affect the physical system \cite{kritzinger2018digital}. In a DT, the connection is automatic, purposeful, and, where an execution infrastructure exists, bidirectional; that is, data from the real system update the model, and the model output can also be returned to the physical system in the form of recommendations, corrections, or control \cite{rasheed2019digital_arxiv,  tao2018digital, thelen2022comprehensivepart1, kritzinger2018digital}. This distinction is important for TDDT because TDDT is not limited to the representation or monitoring of a single domain but seeks the operational connection of multiple domain twins at the levels of state, error, objective, decision, and control.

\subsection{Common Classifications of Digital Twins: Monitoring, Predictive, Prescriptive, and Autonomous}

One common classification of DT divides it based on the level of decision-making function and the degree of automation. In this view, the twin begins at the level of state monitoring, then progresses to future prediction, action recommendation, and finally automatic control or correction \cite{verdouw2021digital, kritzinger2018digital, fuller2020digital, thelen2022comprehensivepart1}. In smart agriculture, types such as Monitoring, Predictive, Prescriptive, and Autonomous have also been used to explain the functional maturity of DT \cite{verdouw2021digital}.

At the Monitoring level, the main objective is to increase observability and reflect the current state of the system \cite{verdouw2021digital, kritzinger2018digital, jones2020characterising}. At the Predictive level, physical, data-driven, or simulation models are used to predict future behavior and assess risk \cite{verdouw2021digital, fuller2020digital, thelen2022comprehensivepart1}. At the Prescriptive level, the twin evaluates intervention options, “what-if” scenarios, constraints, and the cost function, and recommends the appropriate decision \cite{verdouw2021digital, thelen2022comprehensivepart1, tao2018digital}. At the Autonomous level, the output of the twin can affect the physical system automatically or semi-automatically through feedback, control, or reconfiguration \cite{kritzinger2018digital, fuller2020digital, rasheed2020digital}. This classification shows that moving from monitoring to control requires the connection of data, model, feedback, and decision; however, it still does not necessarily explain how several heterogeneous domains should be integrated into a shared state and operationally coupled decision-making.



\subsection{Cross-Domain Approaches in Digital Twins and Their Limitations}

Cross-Domain Digital Twin (CDDT): In this article, the term CDDT is used as an abbreviation for cross-domain approaches in the digital twin literature; approaches whose objective is to identify common features, reusable patterns, general development methods, and interoperability mechanisms among different domains \cite{dalibor2022cross, heindl2022structured}.

In this approach, the main focus is on how the concepts, architectures, data, models, and tools of the digital twin can be generalized from one domain, such as manufacturing, maintenance, product, infrastructure, or cyber–physical systems, to other domains \cite{glaessgen2012digital,kritzinger2018digital,jones2020characterising}. For example, cross-domain studies in software engineering show that DTs in diverse domains, despite their application-specific differences, have common requirements such as modeling, data connectivity, synchronization, simulation, monitoring, and decision support \cite{dalibor2022cross}. Structured cross-domain analyses also show that DT development is a combination of domain-independent and domain-dependent steps; in this sense, part of the architecture, data, and development process can be generic, while an important part of the meaning, model, and decision objective remains dependent on the specific domain \cite{heindl2022structured}. Therefore, in this article, CDDT is not considered a formal standard category, but rather an operational naming for the level of comparison, standardization, reuse, semantic mapping, data exchange, and interoperability among domains \cite{dalibor2022cross, heindl2022structured, david2025interoperability}.

Despite these advantages, CDDT does not inherently and by itself require that the output of one domain be able to operationally change the state, constraint, objective, error, decision, or control of another domain; although, in specific implementations, it can be combined with decision-making or control mechanisms \cite{heindl2022structured, david2025interoperability, jones2020characterising}. Therefore, although the cross-domain approach is necessary for reducing conceptual fragmentation, increasing reusability, and developing general DT frameworks, this article argues that, for the objective intended in this article, namely the operational and time-dependent coupling of data, model, state, error, objective, decision, and control among domains, it is not considered sufficient \cite{jones2020characterising, michael2022integration, vanbeek2023digital}.

In such systems, twins must not only be comparable or reusable, but also be able to be connected in a composite and operational manner in the form of a Digital Twin System-of-Systems \cite{jones2020characterising, michael2022integration}. Therefore, this limitation provides the basis for introducing the concept of the Trans-Domain Digital Twin; a proposed approach in which the connection among domains goes beyond the level of interoperability and reuse and becomes the coupling of data, model, state, error, objective, decision, and control \cite{michael2022integration, vanbeek2023digital, amiri2025jumeaux}.

\begin{figure}[!htbp]
\centering
\includegraphics[width=0.83000\linewidth,alt={Layered cross-domain architecture connecting independent domain twins to interoperability, semantic mapping, reference models and decision-support applications.}]{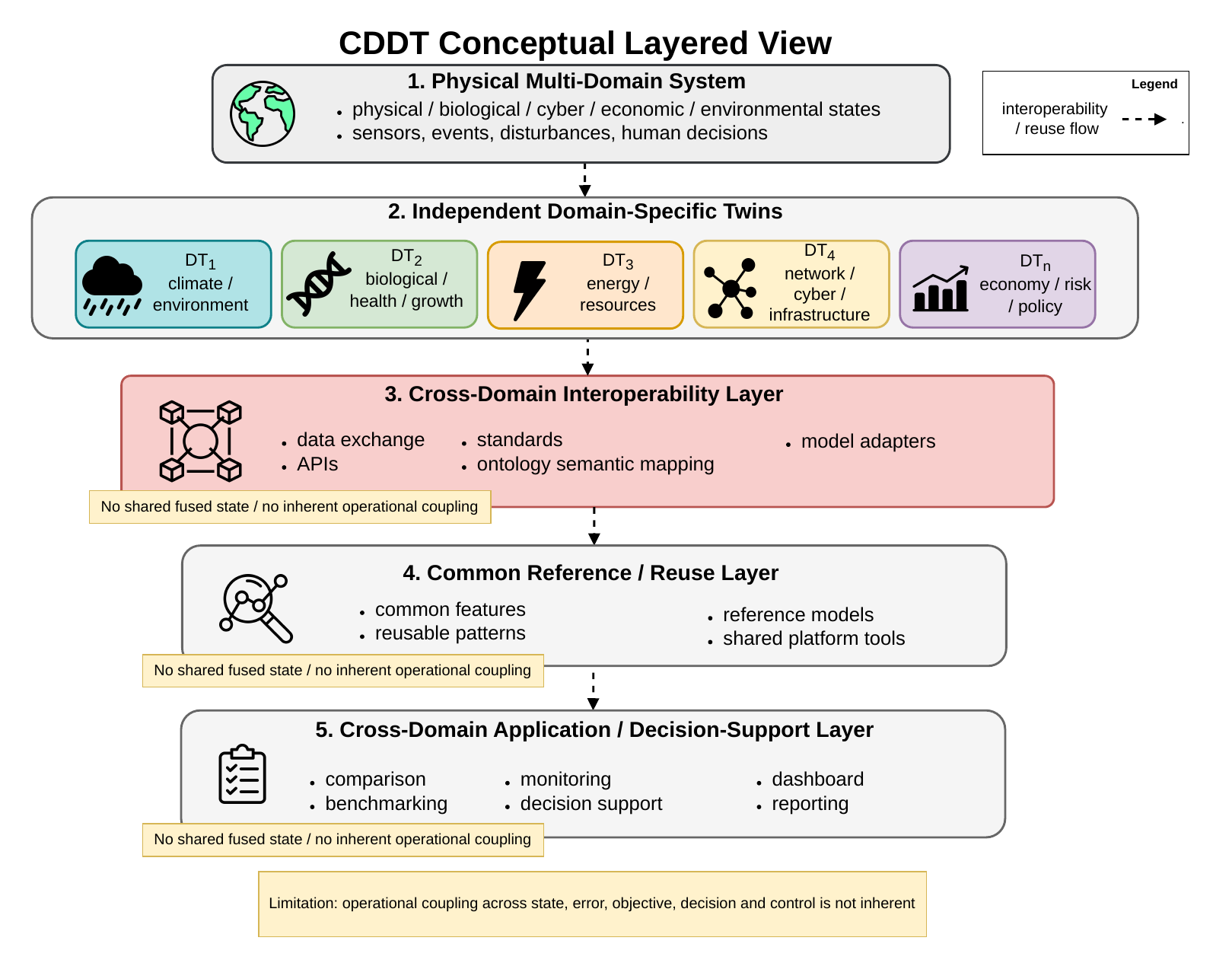}
\caption{Layered Architecture of CDDT - Independent domain-specific twins are connected through data exchange, Application Programming Interfaces (APIs), standards, semantic mapping, and reference models for comparison, monitoring, reporting, and decision support, without inherently creating operational coupling of decision and control among domains.}
\label{fig:cddt-architecture}
\end{figure}

In the inferential architecture of this article for CDDT, analytical or intelligent outputs, including recommendations generated by Artificial Intelligence (AI), are mainly located in the Cross-Domain Application layer in the form of monitoring, dashboard, reporting, and decision support; this level is not equivalent to operationally coupled control among domains. Figure \ref{fig:cddt-architecture} shows the general overview of CDDT.

\section{Definition and Position of the Trans-Domain Digital Twin}
\label{sec:definition}
\subsection{Research Gap in the Operational Connection of Domain Twins}

The digital twin literature covers representation, monitoring, prediction, interoperability, and the composition of multiple independent twins; however, it does not provide an explicit and integrated method for the time-dependent transfer of the effect of one domain to the state, error, objective, constraint, decision, or control of another domain \cite{jones2020characterising, thelen2022comprehensivepart1, dalibor2022cross, michael2022integration, vanbeek2023digital, david2025interoperability}. Cross-Domain approaches mainly focus on comparison, reuse, standardization, and semantic mapping \cite{heindl2022structured, dalibor2022cross, david2025interoperability}, while Composite/Federated architectures and DT System-of-Systems also enable the composition of twins, but do not establish a common requirement for a fused state, operational coupling, multiscale coordination, and cross-domain error feedback \cite{michael2022integration, amiri2025jumeaux, vanbeek2023digital, david2025interoperability, vergara2024federated, khedr2025composition}. Therefore, the main gap is the absence of a framework that simultaneously defines, executes, and evaluates these relationships.

This gap is characterized by four measurable deficiencies: 1) the absence of a shared state for aligning time, units, context, and uncertainty; 2) the absence of explicit coupling among the state, error, objective, and control of the domains; 3) the absence of a mechanism for coordinating heterogeneous temporal loops; and 4) the absence of a traceable pathway for effect transfer and error feedback among the twins \cite{michael2022integration, wu2023comprehensive, amiri2025jumeaux, vanbeek2023digital, david2025interoperability, vergara2024federated, khedr2025composition}.

\subsection{Definition and Positioning of the Trans-Domain Digital Twin}

In this article, TDDT is an architecture at the System-of-Systems (SoS) level that connects multiple heterogeneous DTs through data, models, a shared state, multi-criteria objectives, and a shared decision cycle \cite{kritzinger2018digital, jones2020characterising, thelen2022comprehensivepart1, heindl2022structured, dalibor2022cross, verdouw2021digital, pylianidis2021introducing}. In this architecture, the output of one domain can change the input, state, constraint, risk, objective, or decision of another domain \cite{heindl2022structured, dalibor2022cross, michael2022integration, amiri2025jumeaux, david2025interoperability, vergara2024federated}. Therefore, TDDT is positioned along the continuum of Composite/Federated Digital Twin System-of-Systems, but its distinction lies in the requirement for operational and time-dependent coupling for trans-domain decision-making and control \cite{michael2022integration, amiri2025jumeaux, david2025interoperability, vergara2024federated}.

\subsection{Minimum and Measurable Requirements of TDDT}

A system is considered a TDDT when it includes at least two heterogeneous domain twins, an aligned shared state, at least one coupling at the level of state, error, objective, or control, a traceable cross-domain effect, a temporal synchronization policy, and a feedback pathway for correcting the model or coupling. Data exchange, an application programming interface (API), or semantic mapping, without satisfying these conditions, is not sufficient to classify a system as a TDDT \cite{heindl2022structured, dalibor2022cross, michael2022integration, amiri2025jumeaux, vanbeek2023digital, david2025interoperability}.

\begin{figure}[!htbp]
\centering
\includegraphics[width=0.94769\linewidth,alt={Capability matrix comparing digital twin configurations and identifying the operational capabilities required by TDDT.}]{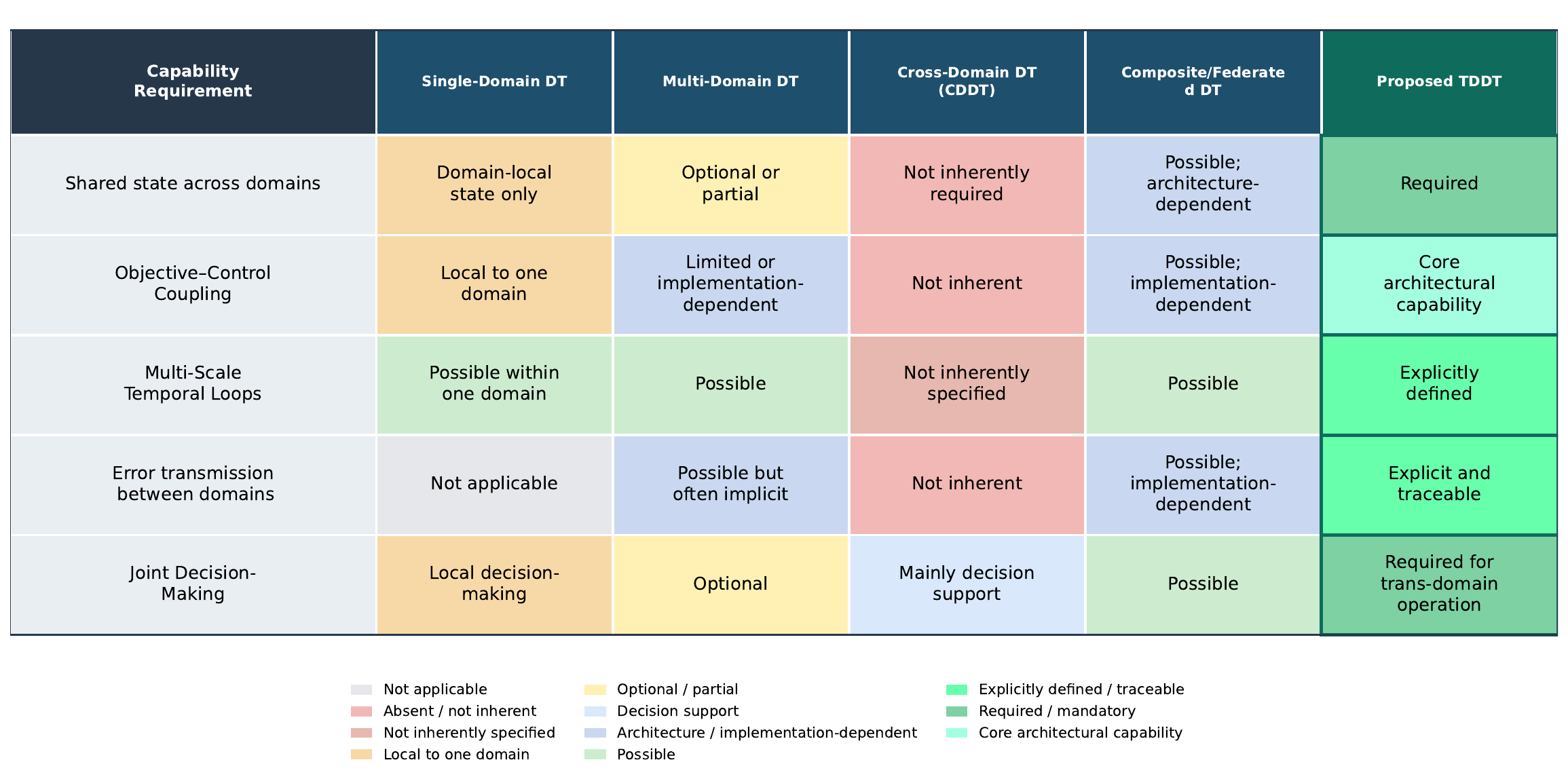}
\caption{Measurable research gap between existing digital twin configurations and the proposed TDDT. This table distinguishes capabilities that are local, optional, implementation-dependent, or explicitly required for trans-domain operations.}
\label{fig:capability-comparison}
\end{figure}

Figure \ref{fig:capability-comparison} summarizes the measurable architectural gap between existing digital twin configurations and the proposed TDDT.

The architecture presented in Section~\ref{sec:architecture} directly addresses this gap; the shared state, operational coupling, temporal coordination, and error feedback are realized in Layers 3.1, 3.2, 3.3, and 7, respectively.

\section{Conceptual Framework and Architecture of TDDT}
\label{sec:architecture}

\begin{figure}[!htbp]
\centering
\includegraphics[width=\linewidth,alt={Seven-layer TDDT architecture showing shared state, coupling, temporal coordination, decision, gated execution and feedback in offline and online modes.}]{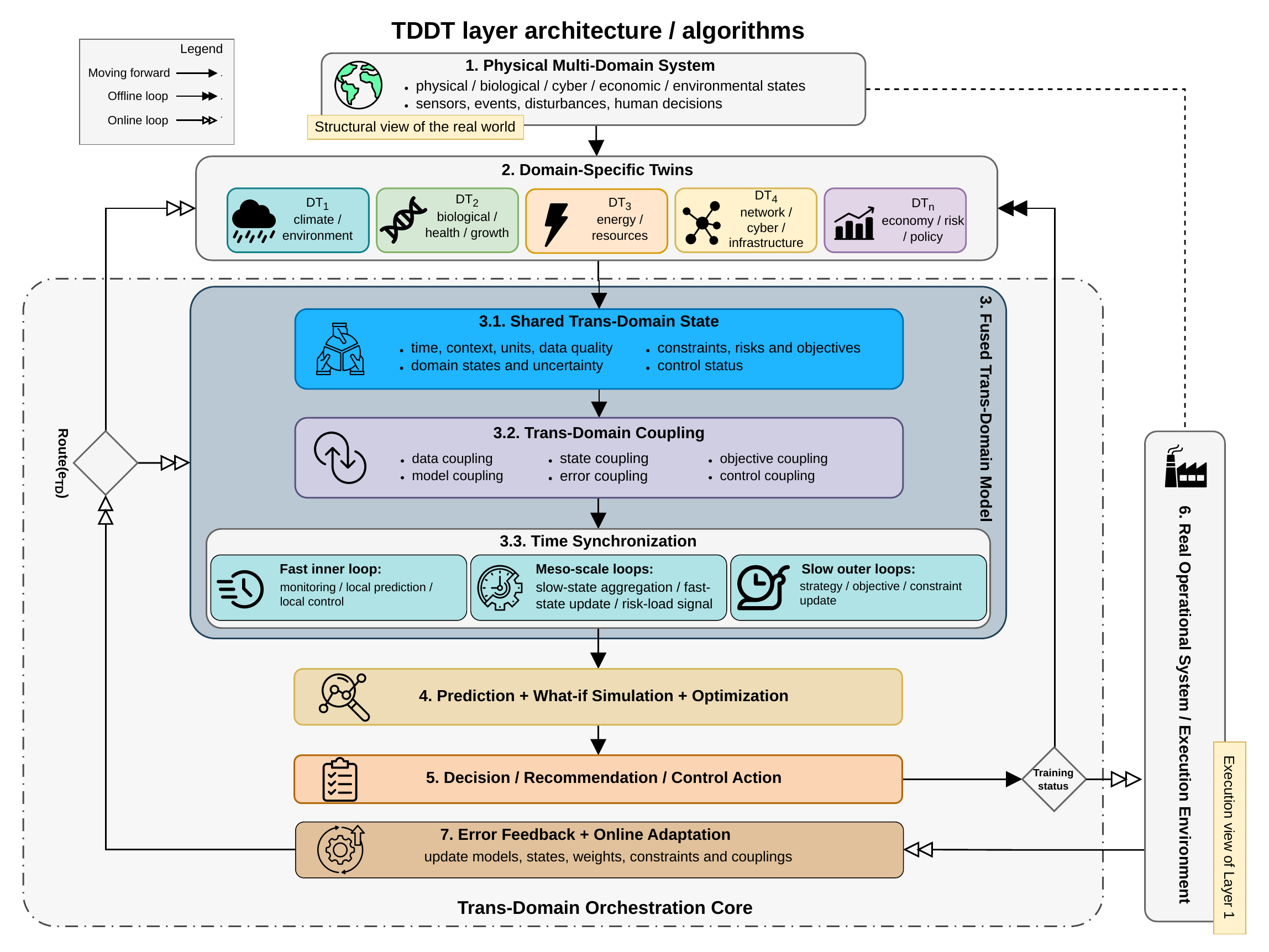}
\caption{Architectural–algorithmic overview of TDDT — Layer 1 shows the structural view of the real multi-domain system, including entities, sensors, actuators, disturbances, and operational constraints; Layer 6 represents its execution view when a decision is applied. In online execution, data and events enter the domain-specific twins from Layer 1; the training-status gate permits Layer 5 actions to reach Layer 6, and Layer 7 routes observed execution errors back for model, state, constraint, and coupling correction. In offline training, the gate blocks the path to Layer 6; training and validation remain within Layers 2–5 without activating execution feedback from Layer 7. Status management is configured in TDOC.}
\label{fig:tddt-architecture}
\end{figure}

The conceptual framework of TDDT includes seven main layers: the multi-domain physical layer, domain-specific twins, the trans-domain shared state, the coupling layer, fast and slow loops, the multi-criteria decision-making layer, and the feedback and online adaptation layer. Figure \ref{fig:tddt-architecture} shows these layers. The position of this framework is defined along the continuum of Digital Twin System-of-Systems and Federated/Composite Digital Twins; however, its distinction in this article lies in establishing operational coupling among domains for trans-domain decision-making and control \cite{khedr2025composition, michael2022integration, david2025interoperability, bolender2021self, amiri2025jumeaux}.

The trans-domain orchestration core, or TDOC, is the central orchestration mechanism of TDDT, which coordinates data exchange, shared state, operational coupling, temporal synchronization, decision-making, and error return among domain-specific twins, the fused model, the decision layer, and online adaptation.

In this architecture, the role of “device communication, data acquisition, timestamping, unit conversion, data quality control, and command transmission to actuators” is considered as an interface sublayer between Layer 1 and Layer 2; this sublayer is functionally mappable to the Device Communication Entity in International Organization for Standardization (ISO) 23247 and to the smart transducer interface layer in Institute of Electrical and Electronics Engineers (IEEE) 1451.

\subsection{Scope, Principles, and Minimum Requirements for TDDT Compliance}

The criteria presented in this section constitute the article’s proposed minimum requirements for distinguishing TDDT from merely multi-domain, Cross-Domain, and Composite/Federated architectures \cite{heindl2022structured, dalibor2022cross, bolender2021self, michael2022integration, david2025interoperability, vergara2024federated, khedr2025composition}. This architecture is based on the relative independence of domain twins, composability, traceable operational effects, temporal coordination, and feedback \cite{bolender2021self, michael2022integration, fuller2020digital, david2025interoperability, khedr2025composition}.

In this formulation, a system is minimally considered a TDDT when it includes at least two heterogeneous twins, an aligned shared or distributed representation, at least one coupling at the level of model, state, error, objective, or control, a traceable cross-domain effect, a temporal coordination policy, multi-domain-based decision-making, and a feedback pathway. This indicates that data exchange, an API, an ontology, or a shared dashboard without operational effects among domains is insufficient \cite{heindl2022structured, dalibor2022cross, michael2022integration, david2025interoperability}.

The seven-layer architecture, TDOC, the meso loop, single-episode training, Context Reference Patterns (CRP), Stage-Aware Reference Guidance (SARG), Meso Reference-Guidance (MRG), and Knowledge Stores (KStores) are the formulations and mechanisms proposed in this article; therefore, their exact implementation is not a general requirement for classification, provided that equivalent functions for the shared state, coupling, coordination, decision-making, and feedback are realized.

\subsection{Multi-Domain Physical Layer}

The first layer includes real entities, subsystems, processes, resources, and actors in the operational environment, such as climate, energy, humans, living organisms, infrastructure, industrial assets, networks, economy, or the natural environment. In the real world, these entities are not independent and affect one another through causal, temporal, spatial, and operational dependencies. In complex systems, information management must cover the stages of design, construction, operation, and maintenance, alongside technical, organizational, data-related, and interoperability dimensions \cite{khedr2025composition, michael2022integration, heindl2022structured, david2025interoperability}. In TDDT, this layer is the main source of data, disturbances, uncertainty, and feedback.

In practical implementation, each sensor and actuator in this layer should be described together with operational metadata such as a unique identifier, quantity type, unit, location, sampling rate, accuracy, measurement range, health status, timestamp, data quality, and uncertainty. This metadata can be maintained as Transducer Electronic Data Sheets (TEDS) or Virtual TEDS according to the logic of IEEE 1451, so that raw data from the real environment can be interpretable, validatable, and synchronizable before entering the domain-specific twins.

\subsection{Domain-Specific Twins and Specialized Sub-Simulators}

The second layer includes domain-specific twins. Each domain has a specialized twin that can be built based on a physical model, data-driven model, agent-based model, simulator, statistical model, machine learning model, or a combination of them; therefore, the construction, encapsulation, and preparation of the domain-specific model or simulator are located in this layer, and its objective is to reflect the state, predict behavior, and generate indicators that can be used by other domains. This idea is consistent with the DT literature, which considers DT as a set of models, data, and technologies connected to the physical world for monitoring, prediction, and decision-making \cite{fuller2020digital, khedr2025composition}. At the System-of-Systems level, each domain-specific twin can initially be developed independently, but it must become composable and interoperable within the higher-level architecture \cite{khedr2025composition, michael2022integration}.

A “sub-simulator” refers to the specialized simulator of a domain or subsystem in Layer 2 and represents only part of the multi-domain system. In the literature, a co-simulator is the framework for scheduling and concurrently executing multiple sub-simulators, such as FMI for the exchange/co-simulation of dynamic models or HLA for distributed simulation. In the TDDT approach, Layer 3 goes beyond co-simulation and transforms the outputs into a shared state, operational coupling, and trans-domain decision-making, which we call the fused trans-domain model \cite{modelica2024fmi302, ieee2025hla1516}.

In this layer, each domain twin has three main tasks: reflecting the state of the domain, predicting the behavior of the domain, and generating indicators that can be used by other domains.

\subsubsection{Fluid dynamics models for representing climate, flow, and environment}

Models from the fluid dynamics family, including CFD\footnote{Computational Fluid Dynamics}, RANS\footnote{Reynolds-Averaged Navier–Stokes}/LES\footnote{Large-Eddy Simulation}, and FFD\footnote{Fast Fluid Dynamics}, are among the most fundamental model families for representing local climate, airflow, heat transfer, humidity, gas, and shear stresses in domains such as closed agricultural halls, aerospace, ventilated buildings, and bioreactors; these models can reveal the spatiotemporal structure of flow and the effects of geometry, boundaries, and actuators \cite{blanesvidal2008application, zuo2010fast, hutmacher2008computational, slotnick2023perspective}. In TDDT, these models can be used in a mechanistic/semi-mechanistic form for accurate simulation, or in the form of FFD, Reduced-Order Models (ROMs), POD\footnote{Proper Orthogonal Decomposition}/DMD\footnote{Dynamic Mode Decomposition}, and data-driven surrogates for fast prediction that can be used in decision loops \cite{blanesvidal2008application, slotnick2023perspective}.

\subsubsection{Lightweighting, analysis, or surrogation}

In this layer, high-fidelity models, such as models derived from CFD, can be transformed into lighter, more analyzable, or fast surrogate models through methods such as ROMs, POD, DMD, or interpolation, due to their high computational and time costs. These methods are mainly located in the second layer, because they are used for order reduction, dynamic analysis, reducing computational cost, and producing faster outputs from domain-specific simulators.

\subsubsection{Encapsulation and standard interfaces}
FMI\footnote{FMI is an open and tool-independent standard for the exchange, integration, and co-simulation of dynamic models, which defines the interface and model packaging format for exchange among different tools \cite{modelica2024fmi302}} provides the outputs of domain-specific models, after being encapsulated in Functional Mock-up Units (FMUs)\footnote{An FMU is an executable/compressed package compliant with the FMI standard, which contains the model, variables, executable functions, the Extensible Markup Language (XML) file describing the model, and, if needed, the binaries or code required for simulation.}, to Layer 3 as exchangeable, time-stamped, unit-aware variables with causal direction; it should be noted that FMI/FMU is mainly used for the encapsulation and exchange of dynamic models and domain-specific simulators. In contrast, the standard connection of real sensors and actuators, metadata reading, calibration, health status, and command transmission to transducers can be performed through the logic of IEEE 1451, including TIM\footnote{Transducer Interface Module}, NCAP\footnote{Network-Capable Application Processor}, and TEDS, in the device communication sublayer. Therefore, in TDDT, FMI for “models” and IEEE 1451 for “real sensors/actuators” play complementary roles. This transition can be part of the twinning process, because it supports the physical–virtual interface for the synchronized representation of the twin with the real system at a specified level of frequency, fidelity, and data quality.

\subsection{Fused Trans-Domain Model}

The third layer includes the fused trans-domain model or state space, which transforms the outputs of domain-specific twins into a shared, coupled, and temporally synchronized progressive state, so that trans-domain prediction, decision-making, and control become possible. This layer is an upstream compositional mechanism that receives the outputs of the twins, simulators, and specialized models of the second layer and reorganizes them into a shared operational state that is couplable, synchronizable, and usable for trans-domain decision-making \cite{khedr2025composition, michael2022integration, david2025interoperability}. Accordingly, in the proposed architecture of this article, the fused state space or fused simulator results from the integration of sublayer 3.1 (sub-section \ref{sublayer_3.1}), sublayer 3.2 (sub-section \ref{sublayer_3.2}) , and sublayer 3.3 (sub-section \ref{sublayer_3.3}). Therefore, the output of these sublayers is transformed into a shared computational–semantic space in which the effect of a change in one domain on other domains can be predicted, evaluated, and controlled \cite{michael2022integration, david2025interoperability, amiri2025jumeaux}. The output of this shared trans-domain state is then transferred to the decision-making, control, and online adaptation layers \cite{david2025interoperability, bolender2021self, amiri2025jumeaux}.

\subsubsection{Ontology, Semantic Mapping, and Conceptual Alignment of Domains}

In this layer, first, the objective and scope of the ontology and the competency questions regarding time, location, unit, data quality, domain state, uncertainty, constraints, and decision are determined according to the requirements of the TDDT architecture \cite{noy2001ontology, gruninger1995methodology}. Then, the vocabulary of each domain, the main concepts, class hierarchies, relations, data properties, the domain and range of relations, logical constraints, and real or simulated instances are extracted \cite{noy2001ontology, fernandezlopez1997methontology}.

To connect multiple domain-specific models or twins, the corresponding components are identified through lexical, unit-based, structural, semantic, temporal, spatial, and causal matching, and the type of mapping or relation between them is determined \cite{uschold1995towards, euzenat2013ontology}. These mappings can be of the type of SKOS\footnote{Simple Knowledge Organization System} relations such as $exactMatch$, $closeMatch$, $broadMatch$, $narrowMatch$, and $relatedMatch$, or of the type of ontological/logical relations such as $equivalentClass$ and $subClassOf$ \cite{euzenat2013ontology, matentzoglu2022simple}. Finally, the mappings are corrected according to differences in unit, level of abstraction, context, or direction of relation, and are stored as a mapping table, shared vocabulary, or bridge ontology, so that the outputs of domain-specific twins can be transferred to the shared trans-domain state and then to the operational coupling layer \cite{euzenat2013ontology, matentzoglu2022simple}.

In addition to mapping domain concepts, it is also necessary to align the metadata of sensors and actuators with the shared TDDT ontology; for example, measurement unit, installation location, sampling rate, accuracy, actuator type, health status, and data quality should be mapped to the corresponding concepts in the shared trans-domain state. This makes the data obtained from smart transducers semantically, unit-wise, temporally, and trust-wise consistent before entering the shared state.

\subsubsection{Trans-Domain Shared State and Data Exchange among Domains} \label{sublayer_3.1}
In Layer 3, Sublayer 3.1 is the point of overlap between CDDT and TDDT and, by using ontology, transforms the important information of each domain into a semantic, temporal, and operational representation of the whole system, termed the shared trans-domain state ($S_{\mathrm{TD}}(t)$; Equation~\eqref{eq:shared_td_state}) so that it can be understandable and usable for other domains. The importance of such a layer is consistent with the literature on multi-domain DT; multi-domain models identify and organize domains such as geometry, structure, data, interaction, application, and time span/horizon in order to create a shared understanding of the system \cite{jiang2023multi}. From the perspective of interoperability as well, heterogeneous twins cannot work together effectively without a shared data model, semantics, protocol, and interface \cite{david2025interoperability}.

\begin{equation}
\label{eq:shared_td_state}
S_{\mathrm{TD}}(t) =
\left\{
\begin{array}{l}
\text{time},\ \text{context},\ \text{device metadata},\\
\text{domain states},\ \text{data quality},\ \text{uncertainty},\\
\text{constraints},\ \text{risks},\ \text{objectives},\\
\text{control status}
\end{array}
\right\}
\end{equation}

The device metadata component includes the sensor/actuator identifier, unit, location, sampling rate, health status, calibration, and timestamp, and acts as a bridge among the physical layer, the device communication sublayer, and the domain-specific twins.

In this sublayer, the output of each domain-specific twin is transformed into the shared trans-domain state through state alignment; that is, the heterogeneous states of the domains are aligned in terms of time, unit, context, quality, uncertainty, constraint, objective, and control status, so that they can be used for trans-domain coupling and decision-making.

\subsubsection{Coupling among Domains—Data, Model, State, Error, Decision, and Control} \label{sublayer_3.2}

In cross-domain DT, the main focus is on extracting common patterns, general development methods, domain-independent steps, and the possibility of reuse across domains \cite{heindl2022structured}. However, in Sublayer 3.2 of Layer 3 of TDDT, the objective is the operational influence of one domain on other domains; that is, the output of one domain should be able to change the input, constraint, objective, error, or decision of another domain.

\begin{figure}[!htbp]
\centering
\includegraphics[width=0.85108\linewidth,alt={Six operational coupling categories from data and model exchange to state, error, objective and control coupling.}]{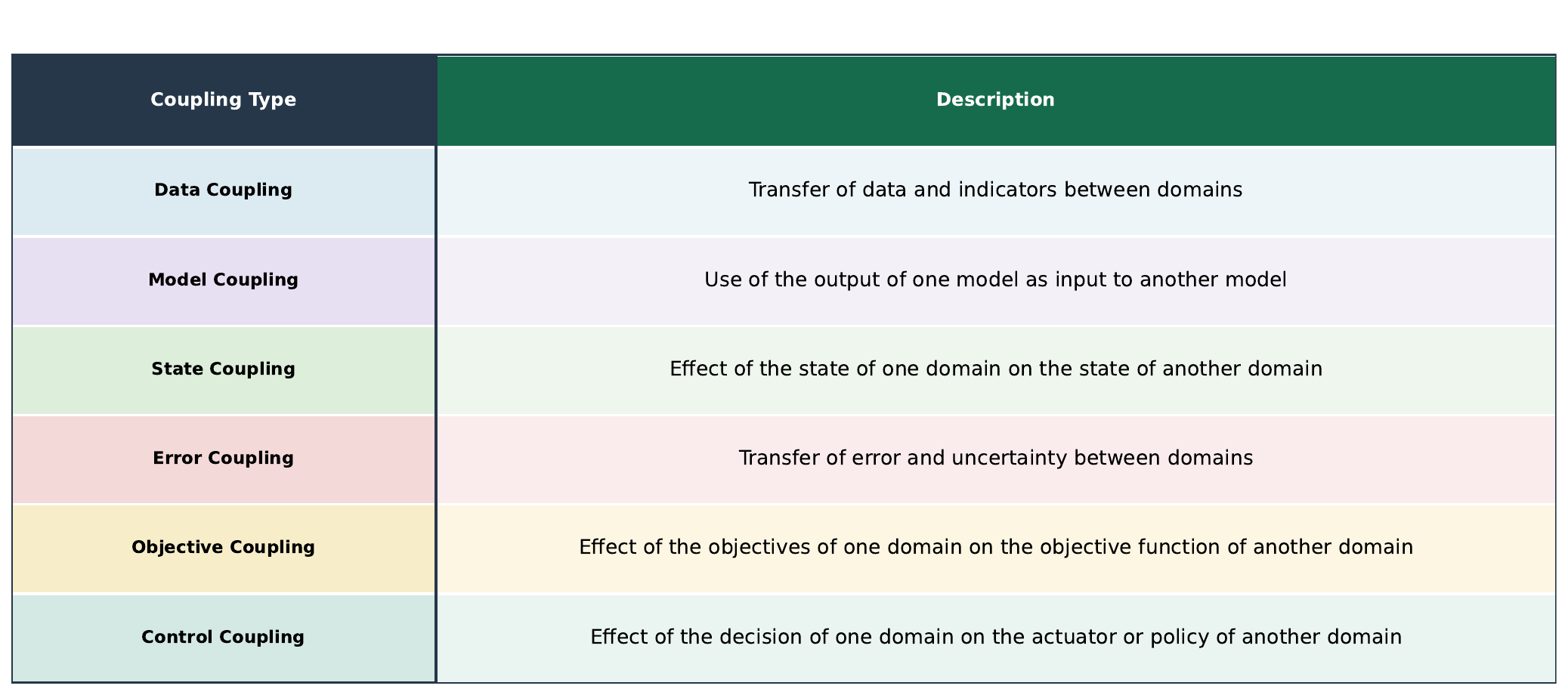}
\caption{Proposed classification of operational coupling in TDDT. Data, model, state, error, objective, and control coupling describe progressively stronger forms of cross-domain influence, ranging from information transfer to direct modification of another domain’s objective, decision, or actuator policy.}
\label{fig:coupling-types}
\end{figure}

In this layer, the proposed types of operational coupling in TDDT are summarized in Figure \ref{fig:coupling-types}.

This layer is consistent with the literature on Digital Twin System-of-Systems; individual twins must be composable and integrable horizontally, vertically, and from different perspectives in order to produce the behavior of the larger system \cite{khedr2025composition, michael2022integration}. At the same time, the precise classification of data, model, state, error, objective, decision, and control couplings is the specific formulation of this article for distinguishing TDDT from cross-domain DT \cite{heindl2022structured, amiri2025jumeaux}.

\subsubsection{Progressive Coupled Time Loops}\label{sublayer_3.3}
In Sublayer 3.3 of Layer 3 of TDDT, operational asynchrony among domains is managed through progressive coupled time loops. Each domain evolves at its own specific speed; domains such as climate, flow, network, energy, or actuators usually require fast inner loops, whereas domains such as growth, health, degradation, economy, resilience, policy, or sustainability change over slower horizons and serve as guides, constraints, or objective-correcting elements for the fast domains. The literature on multi-domain DT, digital twin, and DT systems-of-systems also shows that real-world systems are composed of heterogeneous subsystems, data, models, and behaviors, and operate across different spatial, temporal, and functional scales \cite{fuller2020digital, khedr2025composition, michael2022integration, heindl2022structured, david2025interoperability, bolender2021self, amiri2025jumeaux}. Therefore, in the TDDT architecture, to cover this operational asynchrony, at least two main temporal levels are considered:

\[
\begin{array}{ll}
\textbf{Fast inner loops:} &
\text{monitoring} \rightarrow \text{prediction} \rightarrow \text{local control} \\
&
\Delta t = \text{seconds / minutes} \\[6pt]
\textbf{Slow outer loops:} &
\text{assessment} \rightarrow \text{strategy update} \rightarrow \text{objective correction} \\
&
\Delta T = \text{hours / days / weeks}
\end{array}
\]

In this architecture, the fast loop is responsible for monitoring, stability, and short-term response, whereas the slow loop corrects objectives, constraints, strategies, and decision weights based on long-term consequences. Between these two, an intermediate temporal level titled \textbf{meso-scale} and its corresponding loop titled \textbf{meso-loop} are defined, which receives data from both the fast and slow loops and establishes the operational connection among temporal horizons by aggregating fast states, updating slow states, and generating risk/load signals. For example, in closed livestock farming, the domain of harmful gases is a meso-scale domain, because its production depends on slow biological factors such as feed intake, growth status, and livestock genotype, whereas its momentary control is performed through actuators in the fast loop. Therefore, the meso-scale loop has a secondary temporal level as follows:

\[
\begin{array}{ll}
\textbf{Meso-scale loops:} &
\text{aggregation from slow states} + \text{update from fast states} \\
&
\rightarrow\ \text{intermediate-domain prediction} \\
&
\rightarrow\ \text{constraint / risk / load signal for fast control} \\[4pt]
&
\Delta t_m = \text{minutes / hours}
\end{array}
\]

\begin{figure}[!htbp]
\centering
\includegraphics[width=0.62318\linewidth,alt={Fast, meso and slow loops connected by intermediate guidance and long-term objective updates.}]{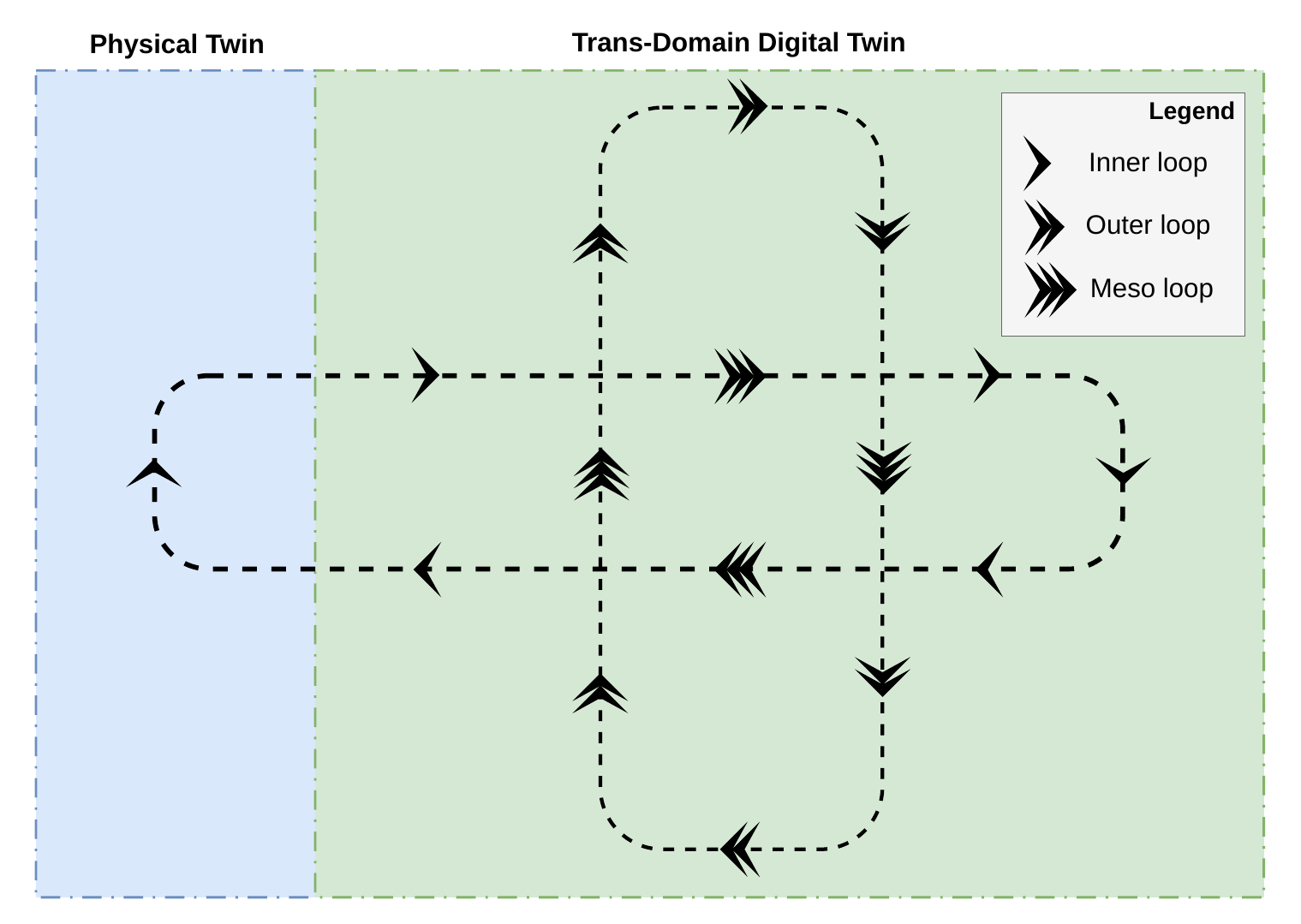}
\caption{Coupled Time Loops in TDDT: The fast loop performs short-term local control; the slow loop updates long-term objectives and constraints; and the meso-loop transfers intermediate-scale data and guidance between these two loops.}
\label{fig:temporal-loops}
\end{figure}

Figure \ref{fig:temporal-loops} shows an overview of these three temporal loops.

\subsubsection{Position of Domain-Specific Model Types in the Fused Model}
In the fused model, the outputs of domain-specific twins and sub-simulators are integrated into a shared computational–semantic trans-domain space; a space that can simultaneously support mechanistic, data-driven, probabilistic, semantic, agent-based, rule-based, and control models. In this layer, ODE\footnote{Ordinary Differential Equation}/PDE\footnote{Partial Differential Equation} models are used for the precise coupling of physical, mechanical, biological, chemical, and transport phenomena; Physics-Informed Neural Networks (PINNs) incorporate physical or biological laws into the learning process; and statistical and data-driven AI / Surrogate / ROM models are used for fast approximation of heavy models and reduction of computational cost \cite{kantaros2025mathematical, raissi2019physics}. In uncertain, incomplete, or time-dependent relationships, Bayesian networks / DBNs\footnote{Dynamic Bayesian Networks} are used to represent probabilistic dependencies, uncertainty, and risk; in conceptual relationships, logical constraints, and inferable decisions, Semantic Reasoning / Rule-Based Models play the role of semantic alignment and application of domain rules; and in micro-level behaviors, ABMs\footnote{Agent-Based Models} agent–agent and agent–environment interactions \cite{orphanou2016dbnextended, okane2021dynamic, horrocks2004swrl, bonabeau2002agent}. In addition, control and decision models, such as control policies, control constraints, setpoint/deadband rules, or predictive control, are represented in Layer 3 as part of Objective/Control Coupling; however, the final selection of the action or executable command is still performed in Layer 5.

\begin{figure}[!htbp]
\centering
\includegraphics[width=0.93154\linewidth,alt={Matrix mapping representative modeling approaches to shared state, coupling, timing and uncertainty functions.}]{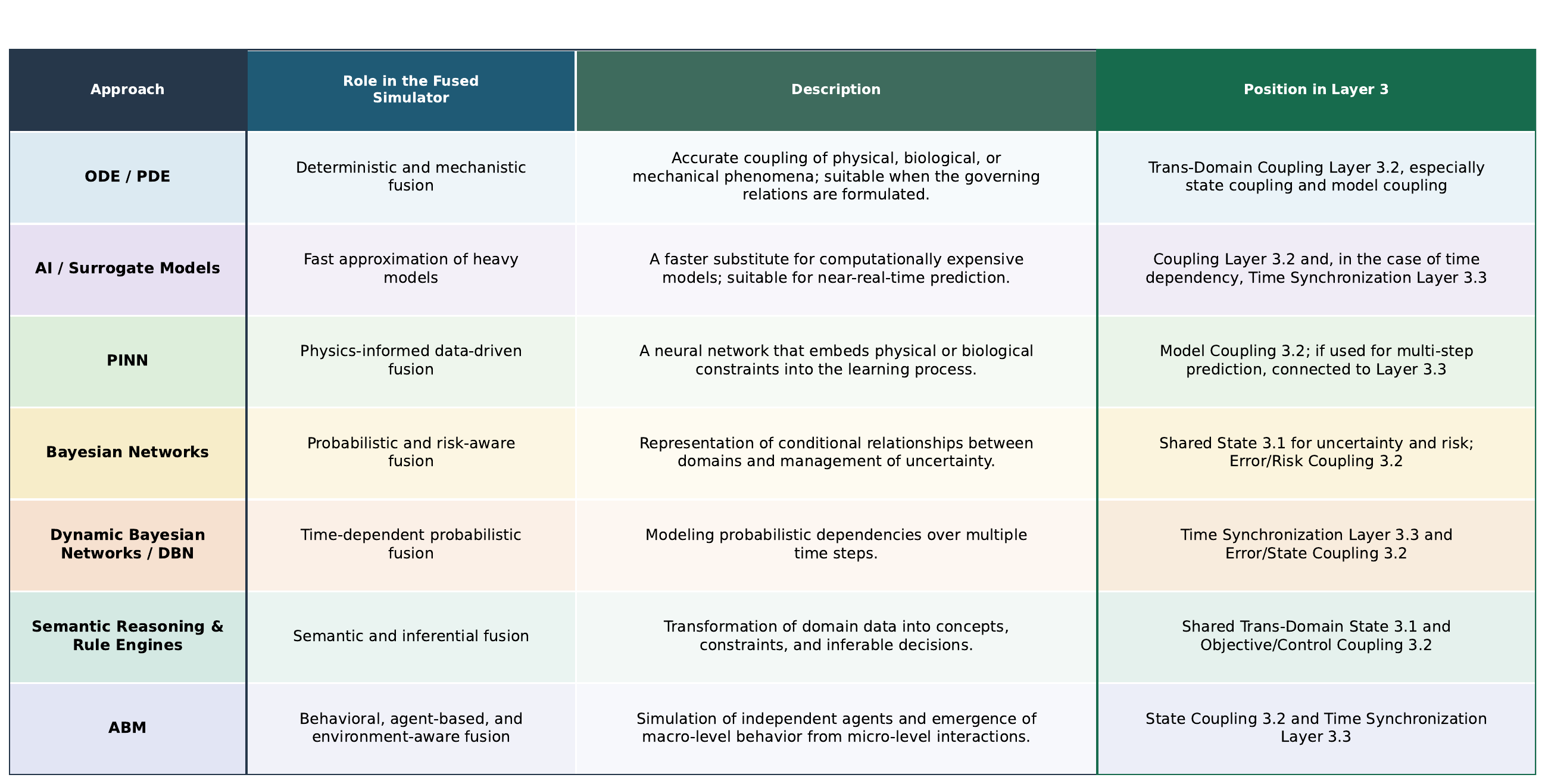}
\caption{Roles and architectural positions of representative modeling approaches in the fused trans-domain model. Mechanistic, surrogate, physics-informed, probabilistic, semantic, and agent-based models support different combinations of shared-state construction, operational coupling, temporal synchronization, and uncertainty representation.}
\label{fig:modeling-approaches}
\end{figure}

Figure \ref{fig:modeling-approaches} summarizes the roles and architectural positions of the main modeling approaches within the fused trans-domain model.

\subsubsection{Concept of Hidden Loop Current and Cyclic Error Signature}

\begin{figure}[!htbp]
\centering
\includegraphics[width=0.67046\linewidth,alt={Direct and indirect cross-domain paths whose differing estimates reveal a hidden loop current.}]{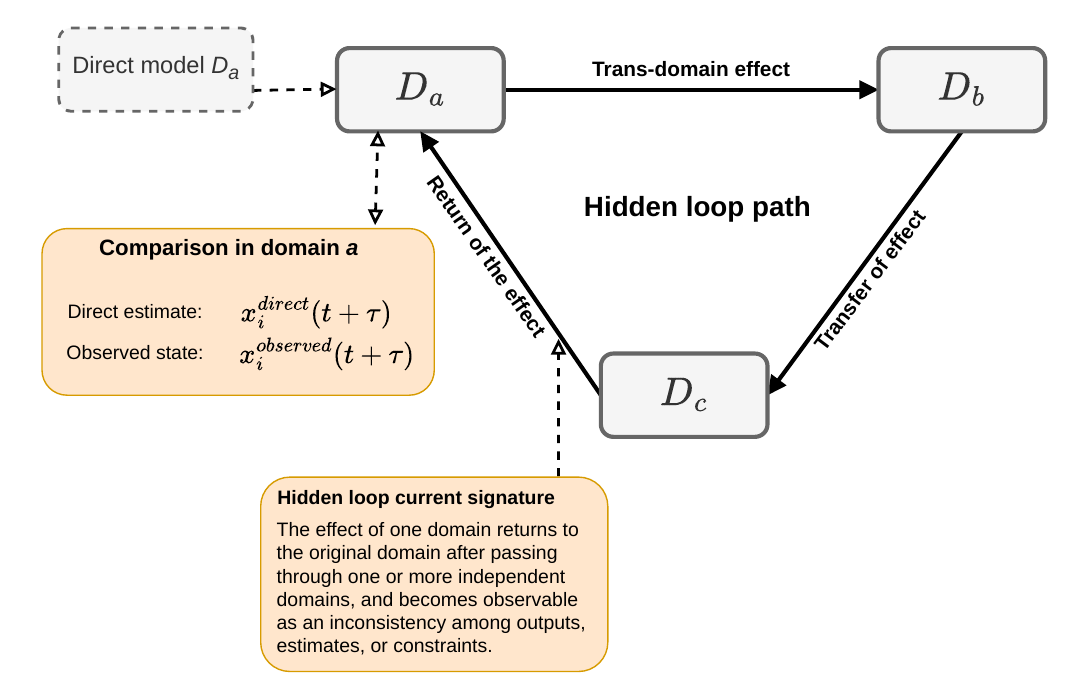}
\caption{Schematic representation of hidden loop current: The effect of domain ($D_a$) reaches domain ($a$) through both the direct path and the hidden path passing through independent domains; the difference between the direct estimate and the observed state activates the hidden loop current index ($LCI$), and is interpreted as a cyclic error signature.}
\label{fig:hidden-loop-current}
\end{figure}

In TDDT, hidden loop current means the indirect and detectable effect of a domain that, after passing through several other domains, returns again to the same domain and creates a behavior that cannot be explained by the direct and linear model of that domain alone. This naming is conceptually inspired by the idea of loop-current and the existence of indirect signatures in physical systems, but in TDDT it is redefined as an architectural indicator for detecting cyclic error and hidden couplings \cite{suetsugu2026microscopic}. In TDDT, this concept is used to identify hidden couplings, returned error, delayed effects, regime shift, and risks that are not evident in the output of a single domain alone, but after passing through paths such as ($D_i \rightarrow D_j \rightarrow D_k \rightarrow D_i$), appear as a persistent difference between direct prediction and observed behavior (Figure \ref{fig:hidden-loop-current}). This idea considers hidden loop current not as a direct variable, but as an indirect, persistent, and multi-domain signature.

The simple formula of the hidden Loop Current Index (LCI) is as follows:

$$
LCI_i(t,\tau)=\left|x_i^{observed}(t+\tau)-x_i^{direct}(t+\tau)\right|,\qquad
LCI_i>\theta_i
$$

where ($LCI_i$) is the \textbf{loop index,} ($x_i^{observed}$) is the \textbf{observed state}, ($x_i^{direct}$) is the \textbf{direct prediction,} ($\tau$) is the \textbf{delay horizon}, and ($\theta_i$) is the \textbf{error threshold}, The application of this index is to indicate whether the error of a domain, after passing through several other domains, has returned again to the same domain and changed the decision, control, or uncertainty. If this index is activated, the feedback path can be transferred to Layer 2 for correcting the domain-specific model and/or to Layer 3 for correcting alignment, coupling, and synchronization.

In terms of the minimum number of domains, TDDT can conceptually be defined with \textbf{at least two heterogeneous domains}; at minimum, there must be one operational relationship between two domain-specific twins. However, for the effective detection of \textbf{hidden loop current}, the presence of \textbf{three domains or three independent estimation paths} is recommended, so that the returned effect through the indirect path can be compared with the direct prediction. The existence of models that examine a domain from different perspectives is not mandatory for the general definition of TDDT; however, in high-risk or low-sensor domains, having multiple independent models of the same domain, such as physical, data-driven, statistical, or mirror models, is strongly recommended for separating domain-specific error from coupling error.

For example, Unmanned Aerial Vehicle (UAV) navigation in Global Navigation Satellite System (GNSS)-independent environments, due to high risk, limited sensors, and returned errors among inertia, wind, altitude, energy, control, and map, shows that in sensitive applications, using multiple domains, sometimes up to 10 to 12 complementary domains, can make the detection of hidden loop current more reliable.

\subsection{Prediction, “What-If” Simulation, and Optimization Operations}

Layer 4 uses the output of the fused trans-domain model for future prediction, analysis of \textbf{what-if}  scenarios, and evaluation of optimization options. By “what-if”, this article refers to testing interventions, constraints, disturbances, objectives, or alternative decisions and comparing their consequences across multiple domains. This role is consistent with the general definition of DT, which considers DT as a tool for monitoring, prediction, analysis, control, and decision-making \cite{fuller2020digital, modelica2024fmi302}. However, in TDDT, prediction is performed in order to evaluate the simultaneous effect of multi-domain state, risk, uncertainty, and constraints; an issue that is also consistent with the literature on Composite DTs / Systems-of-Systems and multi-domain models of complex infrastructures \cite{khedr2025composition, jiang2023multi}. When the execution of multi-model scenarios and co-simulation is required, standards such as FMI can also be used for model exchange, co-simulation, and scheduled execution of models \cite{modelica2024fmi302}.

In this layer, to reduce computational cost, generate fast scenarios, and estimate uncertainty, models such as surrogate models, reduced-order models, simplified physics-based models \cite{raissi2019physics}, PINN, Gaussian Process emulator \cite{maatouk2017gaussian}, or approximate generative models can be used for reducing computational cost, generating fast scenarios, and estimating uncertainty. If the Gaussian model is used for uncertainty prediction, output-error modeling, or generating a fast approximation of scenarios, its main position is in this layer; however, if it is used for error correction with real data, it is transferred to Layer 7 in this architecture. Genetic algorithms \cite{goldberg1989genetic} are also mainly located in this layer for searching scenarios and finding the optimal combination of variables/parameters.

\subsection{Decision, Recommendation, or Control Action}

Layer 5 is the layer that transforms the output of prediction and scenario evaluation into an operational decision, managerial recommendation, control policy, or real-time atomic command; that is, it generates the smallest executable action on an actuator, setpoint, or operational policy and sends it to the next layer. In this layer, the decision function simultaneously evaluates multi-domain error, risk, energy, health, cost, quality, sustainability, and uncertainty. This role is consistent with the definition of DT as a tool for representing, monitoring, predicting, controlling, and optimizing cyber–physical systems \cite{fuller2020digital, bolender2021self}, and in TDDT it is extended to the level of multi-domain decision-making \cite{khedr2025composition, jiang2023multi}.

Since the fused trans-domain model provides the decision layer with the shared state, constraints, uncertainty, and mutual effects of domains, decision-making in TDDT can rely on model-based predictive control such as Model Predictive Control (MPC), instead of blindly searching among options, in order to select constrained and executable actions over a future time horizon \cite{khedr2025composition, camacho2007model}.

Based on the current state, prediction of system behavior over the future horizon, operational constraints, and the multi-criteria objective function, MPC computes an optimal control sequence; usually, its first command or executable action is sent to the execution layer as an atomic command, and the process is repeated in the next step with new data. Therefore, in the TDDT architecture, MPC \cite{camacho2007model} can transform the multi-domain predictions of Layer 4 into a decision, setpoint, or actuator command \cite{fuller2020digital, bolender2021self}.

\subsection{Real Executable System and Decision Application Environment}

Layer 6 is the execution view of the real system in TDDT; that is, the part of the multi-domain physical world on which the output of the decision-making layer is applied, and whose real response is returned to Layer 7 for error calculation and online adaptation. The difference between this layer and Layer 1 is that Layer 1 describes the general structure of physical, biological, cyber, economic, and environmental entities as the source of data and the operational context; whereas Layer 6 represents the execution view of the same system at the time of decision application, that is, the place where commands, recommendations, or control actions are actually executed. This role is consistent with the general definition of DT, which emphasizes the bidirectional connection between the physical entity and the digital model/representation \cite{fuller2020digital}, and, in cyber–physical systems, also includes the control and optimization of the behavior of the real system \cite{bolender2021self}.

In this layer, real data on disturbances, execution constraints, actuator error, human behavior, and decision consequences appear; this necessity is consistent with the literature on Composite DTs / Systems-of-Systems, the challenges of DT integration, and multi-domain models of complex infrastructures \cite{khedr2025composition, michael2022integration, jiang2023multi}.

\subsection{Error Feedback, Online Tuning, and Automatic Adaptation}

Layer 7 calculates the error between the predicted output and the real behavior of the system and uses it to update models, states, weights, constraints, parameters, confidence level, uncertainty, and, if needed, the coupling structure; therefore, uncertainty adaptation is performed in this layer, and its output is returned to Layers 2 and 3 so that the next prediction and decision become consistent with real conditions. This is the layer of \textbf{feedback and online adaptation on offline-learned knowledge,} because a DT gains higher operational value when it adapts to new data, the real state, and changing conditions; an issue that is consistent with the definition of DT as a bidirectional connection between the physical system and the digital representation \cite{fuller2020digital}, as well as with the role of DT in the control, prediction, optimization, and self-adaptation of cyber–physical systems \cite{bolender2021self, amiri2025jumeaux}.

Control or corrective feedback, before being applied to the real system, must pass through the device communication sublayer so that the command can be checked in terms of actuator identifier, allowable range, unit, application time, actuator health status, receipt acknowledgment, and safety constraints. Therefore, the online adaptation of TDDT is performed at the conceptual level of models, but its physical execution depends on the standard sensor/actuator interface and the operational gateway.

If the Gaussian model is used for online tuning, discrepancy estimation, residual correction, error calibration, or parameter updating with real data, its main position is in Layer 7. In this case, the Gaussian output is not directly a control command; rather, in this architecture, it is a correction for the model, weight, constraint, parameter, or uncertainty, which returns through the feedback path to Layer 2 and Layer 3 and then affects the subsequent decisions of Layers 4 and 5 \cite{maatouk2017gaussian, kennedy2001bayesian}.

\subsection{Single-Episode Offline Training and Transformation of Trans-Domain Experience}

In the TDDT framework, single-episode offline training is defined as a continuous process in which domain-specific simulators, the fused model with appropriate fidelity, the predictive decision-maker, and the memory layer are executed together within a single time horizon. The purpose of this training is to provide trans-domain prior knowledge to TDDT before online execution; in such a way that the interaction among state, decision, cost, constraint, and domain feedback is transformed into experience usable by progressive temporal coupling, and the need for costly and high-risk online learning, especially in biological, health, and military applications, is reduced. This logic is aligned with the idea of decision-driven digital twins, predictive control, and multi-domain architectures \cite{yang2019adaptive, shin2024applicability, jiang2023multi}. In this process, MPC is used as the predictive decision-making layer for selecting control actions under constraints and uncertainty \cite{yang2019adaptive}.

\subsubsection{Preparation of CRP and SARG Reference Datasets}

Data preparation in TDDT is performed through two complementary pathways. In the CRP pathway, historical, simulated, or operational data related to the context of each scenario—such as environmental conditions, system state, mission phase, individual characteristics, resources, disturbances, and sensor quality—are collected, cleaned, unit-harmonized, time-synchronized, and resampled at the appropriate temporal scale. Then, for each interval, the contextual feature vector ($x_c$) is extracted and, after normalization, similar samples are classified into Context Reference Patterns using clustering or similarity-based methods: ($\mathrm{CRP}={C_k=(\mu_k,\Sigma_k,n_k,q_k)}_{k=1}^{K}$), where ($\mu_k$), ($\Sigma_k$), ($n_k$), and ($q_k$) respectively represent the context, dispersion, number of samples, and cluster quality. Therefore, in UAV navigation, CRP can represent wind regimes, topography, sensor quality, and flight phase; and in personalized treatment, it can represent phenotype, disease stage, patient status, and history of response to treatment.

In the SARG pathway, the outputs of valid scenarios, optimized simulations, expert opinions, or past successful decisions are divided based on temporal or operational stages, and for each stage, the desired behavior, safe limits, objectives, constraints, and decision weights are extracted. Each guidance pattern can be defined as ($G_s=(z_s^{*},a_s^{*},J_s^{*},\mathcal{C}*s,w_s,\gamma_s)$), which respectively indicates the desired state, reference action, optimal cost or performance, constraints, objective weights, and confidence level of stage (s); as a result, ($\mathrm{SARG}={G_s}*{s=1}^{S}$) forms the staged desired pathway. During training or execution, the current context is mapped to the nearest CRP cluster, and the corresponding SARG is transferred to the optimizer as a prior, guidance, or weight adjustment; therefore, CRP answers “what type of context is the system currently in?” and SARG specifies “what is the desired and optimal behavior in this stage and context?”.

\subsubsection{Generation, Aggregation, and Transfer of Knowledge in Progressive Single-Episode Loops}

At each step of the fast loop, the domain-specific twins and the fused model predict the current state, and the model-based decision-maker selects the appropriate action under the existing constraints and objectives. The result of each step is stored in the step-stream memory as a temporal record, including the trans-domain shared state, decision, action, cost or reward, active constraints, uncertainty, guidance, and decision weights (Equation~\eqref{eq:fast_episode_record}):

\begin{equation}
\label{eq:fast_episode_record}
e_{t}^{\mathrm{fast}}
=
\left(
S_{\mathrm{TD}}(t),
\hat{S}_{\mathrm{TD}}(t+1),
a_t,
J_t,
\mathcal{C}_t,
U_t,
g_t,
w_t
\right)
\end{equation}

At the end of each operational period, the set of fast records is aggregated in the slow loop in order to extract long-term trends, the cumulative response of domains, resource consumption, changes in constraints, state coverage, and decision quality. The output of this stage is a compact episodic record that is stored in the single-episode trans-domain knowledge repository. This record preserves the relationship among context, state, decision, outcome, uncertainty, and the mutual effects of domains, and transforms short-period knowledge into experience usable over longer time horizons; then, the guidance extracted from the slow loop is returned to the fast loop to correct the decisions of the next cycle (Equation~\eqref{eq:episode_aggregation}):

\begin{equation}
\label{eq:episode_aggregation}
e_{d}^{\mathrm{episode}}
=
\operatorname{Aggregate}
\left(
\left\{ e_{t}^{\mathrm{fast}} \right\}_{t \in d}
\right)
\end{equation}

In the meso loop, the contextual features of the aggregated record are first matched with CRP clusters in order to determine the corresponding Context Reference Pattern; then, according to the current stage and the detected context, the corresponding desired pattern is selected from SARG. The difference between the observed or simulated behavior and the desired pattern, together with constraints, uncertainty, and matching quality, is transformed into an inter-loop guidance signal:

\begin{equation}
\label{eq:meso_guidance_mrg}
g_{d}^{\mathrm{meso}}
=
\operatorname{MRG}
\left(
e_{d}^{\mathrm{episode}},
C_{k}^{\mathrm{CRP}},
G_{s}^{\mathrm{SARG}}
\right)
\end{equation}

\begin{figure}[!htbp]
\centering
\includegraphics[width=0.98708\linewidth,alt={Fast records aggregate into episodic knowledge and combine context and stage reference patterns to generate meso guidance.}]{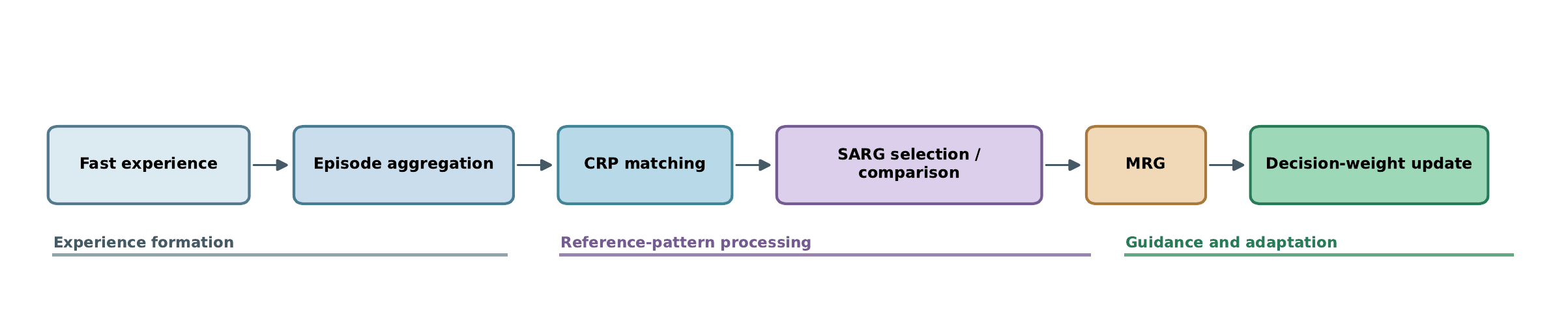}
\caption{Knowledge-transformation pathway in the proposed single-episode TDDT training process. Fast-loop experiences are aggregated into episodic knowledge, matched with CRP, compared with SARG, and compressed into MRG for updating decision weights in the subsequent cycle.}
\label{fig:episode-knowledge-flow}
\end{figure}

In Equation~\eqref{eq:meso_guidance_mrg}, MRG is the inter-loop guidance process or signal that compresses the result of matching the current context with CRP and comparing the stage-wise behavior with SARG. This signal is returned to the fast loop in the next cycle to correct the objectives, decision weights, constraints, and optimizer parameters. The knowledge-transformation pathway is summarized in Figure \ref{fig:episode-knowledge-flow}.

\begin{figure}[!htbp]
\centering
\includegraphics[width=0.83928\linewidth,alt={Progressive coupled learning loops connecting fast experience storage, slow aggregation and guidance returned to the controller.}]{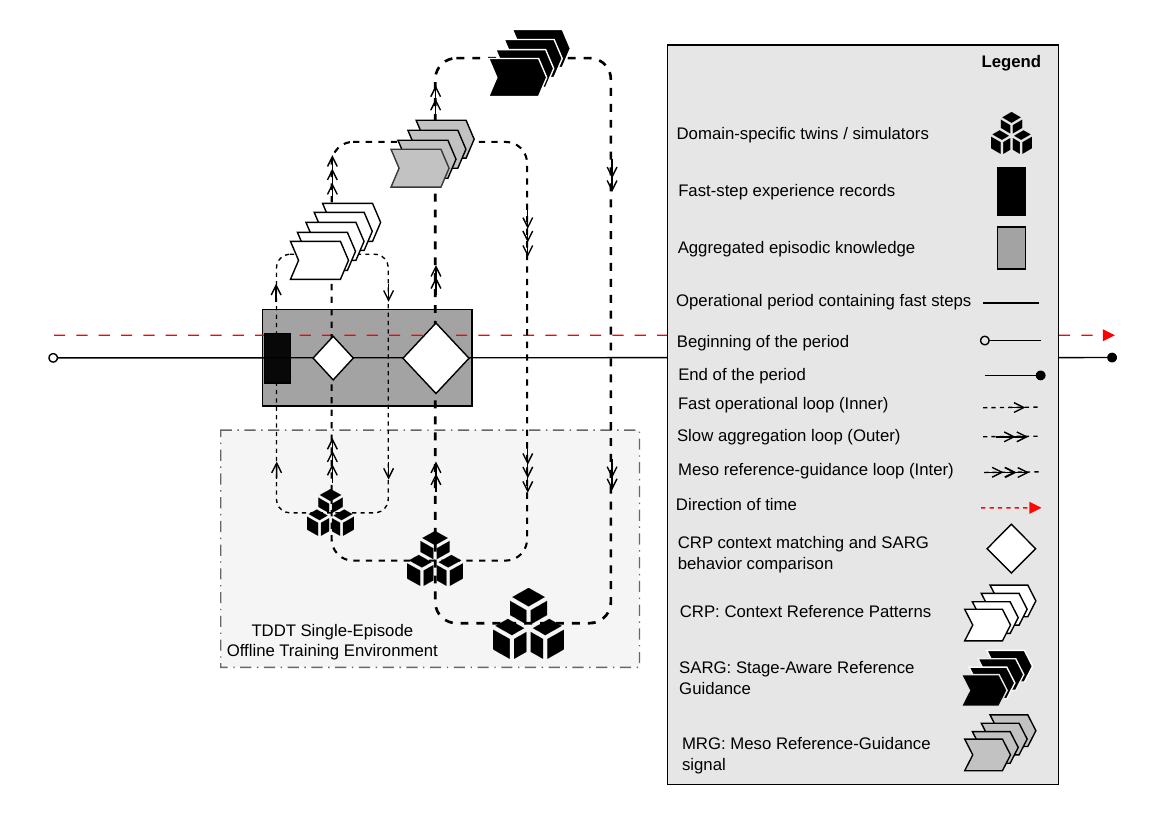}
\caption{Simplified representation of single-episode learning in the progressive coupled loops of TDDT: At each fast step, state, decision, cost, guidance, and decision weights are stored as short-period knowledge; then this knowledge is aggregated at a slower scale, compared with CRP and SARG, and its result is returned to the optimizer and controller as compact guidance.}
\label{fig:progressive-loops}
\end{figure}

\begin{figure}[!htbp]
\centering
\includegraphics[width=0.70246\linewidth,alt={Processing blocks for step-stream memory, episodic knowledge and context-dependent meso guidance.}]{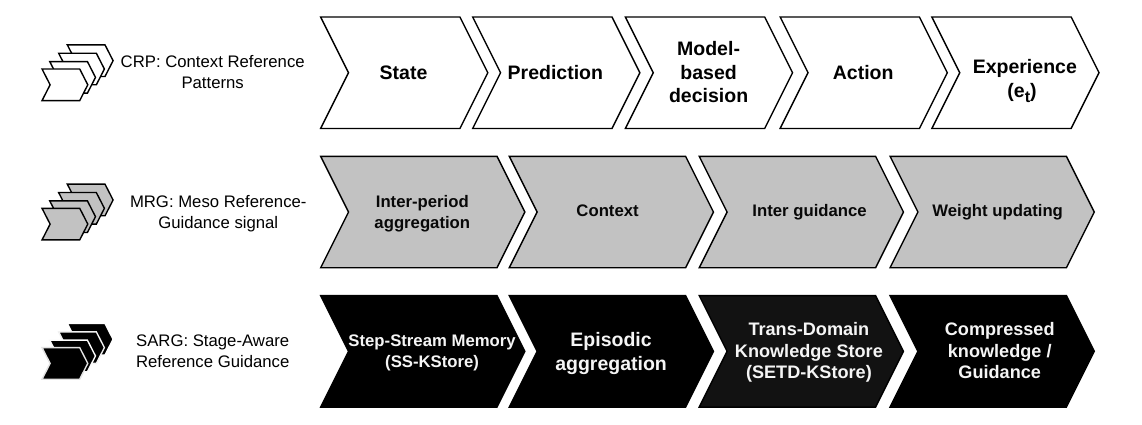}
\caption{Data-processing details in the progressive TDDT loops: the fast path generates the sequence of state prediction, model-based decision, action, and experience for each step and stores it in SS-KStore; the slow path transforms step-by-step experiences into cumulative knowledge in SETD-KStore; and the meso path uses CRP, SARG, and MRG to update the context, inter-loop guidance, and decision weights for the next cycle.}

\label{fig:loop-processing}
\end{figure}

At the end of the episode, the fast records, cumulative knowledge, CRP cluster identifier, selected SARG pattern, MRG guidance, and final weights are stored in the trans-domain knowledge package so that the online controller begins its operation from a valid prior knowledge base, not from zero-shot learning. Figure \ref{fig:progressive-loops} shows the temporal and simplified view of this circulation, and Figure \ref{fig:loop-processing}  describes the stages of data processing and transformation across the fast, meso, and slow loops.

Along this pathway, the Step-Stream Memory Knowledge Store (SS-KStore) serves as the streaming memory for the step-by-step experiences of the fast loop during training, whereas the Single-Episode Trans-Domain Knowledge Store (SETD-KStore) transforms these experiences, after episodic aggregation, into a compact, trans-domain, and loadable knowledge package for online control.

\section{General Formalism and Operational Logic of TDDT}
\label{sec:formalism}

\subsection{General TDDT Formalism}

The following formalism provides a minimal representation of the architecture proposed in this article \cite{bolender2021self, michael2022integration, amiri2025jumeaux, david2025interoperability, vergara2024federated, khedr2025composition}:

\begin{equation}
\label{eq:tddt_formal_tuple}
\mathcal{T}_{\mathrm{TD}}
=
\left\langle
\mathcal{D},
S_{\mathrm{TD}},
\mathcal{C},
\mathcal{L},
\mathcal{J},
\mathcal{U},
\mathcal{K},
\mathcal{F}
\right\rangle,
\qquad
\mathcal{D} = \{D_i\}_{i=1}^{n},\quad n \geq 2
\end{equation}

In Equation~\eqref{eq:tddt_formal_tuple}, ($\mathcal{D}$) denotes the set of domain twins, ($S_{TD}$) the aligned shared or distributed state, ($\mathcal{C}$) the couplings, ($\mathcal{L}$) the temporal policy, ($\mathcal{J}$) the objectives and constraints, ($\mathcal{U}$) the decision, ($\mathcal{K}$) the optional knowledge, and ($\mathcal{F}$) the feedback. Equations~\eqref{eq:shared_td_state_aggregation}, \eqref{eq:domain_coupling_signal}, \eqref{eq:td_optimal_control_feasible}, \eqref{eq:td_error}, and \eqref{eq:td_knowledge_update} specify the corresponding operations.

\begin{equation}
\label{eq:shared_td_state_aggregation}
S_{\mathrm{TD}}(t)
=
\mathcal{A}
\left(
x_1(t_1), \ldots, x_n(t_n), m, q, \Sigma
\right)
\end{equation}

\begin{equation}
\label{eq:domain_coupling_signal}
z_j(t)
=
\sum_{i \neq j}
C_{ij}
\left(
y_i(t-\tau_{ij}),
S_{\mathrm{TD}}(t)
\right)
\end{equation}

\begin{equation}
\label{eq:td_optimal_control_feasible}
u_{\mathrm{TD}}^{*}(t)
=
\arg\min_{u \in \mathcal{U}_{\mathrm{feasible}}}
\left[
\sum_{i=1}^{n}
w_i J_i\left(S_{\mathrm{TD}}, u\right)
+
\lambda_R R
+
\lambda_U U
\right]
\end{equation}

\begin{equation}
\label{eq:td_error}
e_{\mathrm{TD}}(t)
=
y_{\mathrm{real}}(t)
-
\hat{y}_{\mathrm{TD}}(t)
\end{equation}

\begin{equation}
\label{eq:td_knowledge_update}
\Theta_{\mathrm{TD}}(t+1)
=
\mathcal{H}
\left(
\Theta_{\mathrm{TD}}(t),
e_{\mathrm{TD}}(t),
S_{\mathrm{TD}}(t),
U_{\mathrm{TD}}(t),
\operatorname{Route}(e_{\mathrm{TD}})
\right)
\end{equation}

A system is compatible with this formalism only when it includes at least two heterogeneous twins, a traceable operational coupling, an aligned state, multi-domain-based decision-making, and a feedback pathway. The model type, optimizer, number of loops, and memory mechanism are application-dependent.

\subsection{Trans-Domain Objective Function and Multi-Criteria Decision-Making}

In TDDT, due to the transfer of risk or cost from one domain to other domains, the final decision should not be made based on the local optimization of a single domain. Therefore, the trans-domain objective function must simultaneously consider multiple objectives such as performance, safety, energy, sustainability, cost, health, risk, uncertainty, and control quality; this logic is consistent with the role of DT in optimization, decision-making, and uncertainty management \cite{thelen2022comprehensivepart2, rasheed2020digital}. In general, the TDDT objective function can be represented as follows:

\begin{equation}
\label{eq:tddt_objective}
\begin{aligned}
J_{TD} ={}&
w_{1}J_{physical}
+ w_{2}J_{biological}
+ w_{3}J_{energy} \\
&+ w_{4}J_{cyber}
+ w_{5}J_{economic}
+ w_{6}J_{environmental} \\
&+ w_{7}J_{risk}
+ w_{8}J_{uncertainty}
\end{aligned}
\end{equation}

In Equation~\eqref{eq:tddt_objective}, each (J) represents the cost, error, risk, or objective in a domain, and each (w) indicates the decision weight of that domain. A trans-domain decision is selected when it produces the lowest total cost or the highest total utility while satisfying multi-domain constraints (Equation~\eqref{eq:td_optimal_control}):

\begin{equation}
\label{eq:td_optimal_control}
u_{TD} = \arg\min_{u_{TD}} J_{TD}
\end{equation}

Therefore, TDDT goes beyond the level of prediction and becomes a decision-support or control-support mechanism that compares different scenarios based on the shared consequences of multiple domains \cite{verdouw2021digital, thelen2022comprehensivepart2, rasheed2020digital}.

\subsection{Uncertainty and Model Correction in Offline Training}

\begin{figure}[!htbp]
\centering
\includegraphics[width=0.85908\linewidth,alt={Offline simulator and mirror-model comparisons produce residuals and a compact dataset for uncertainty estimation and error routing.}]{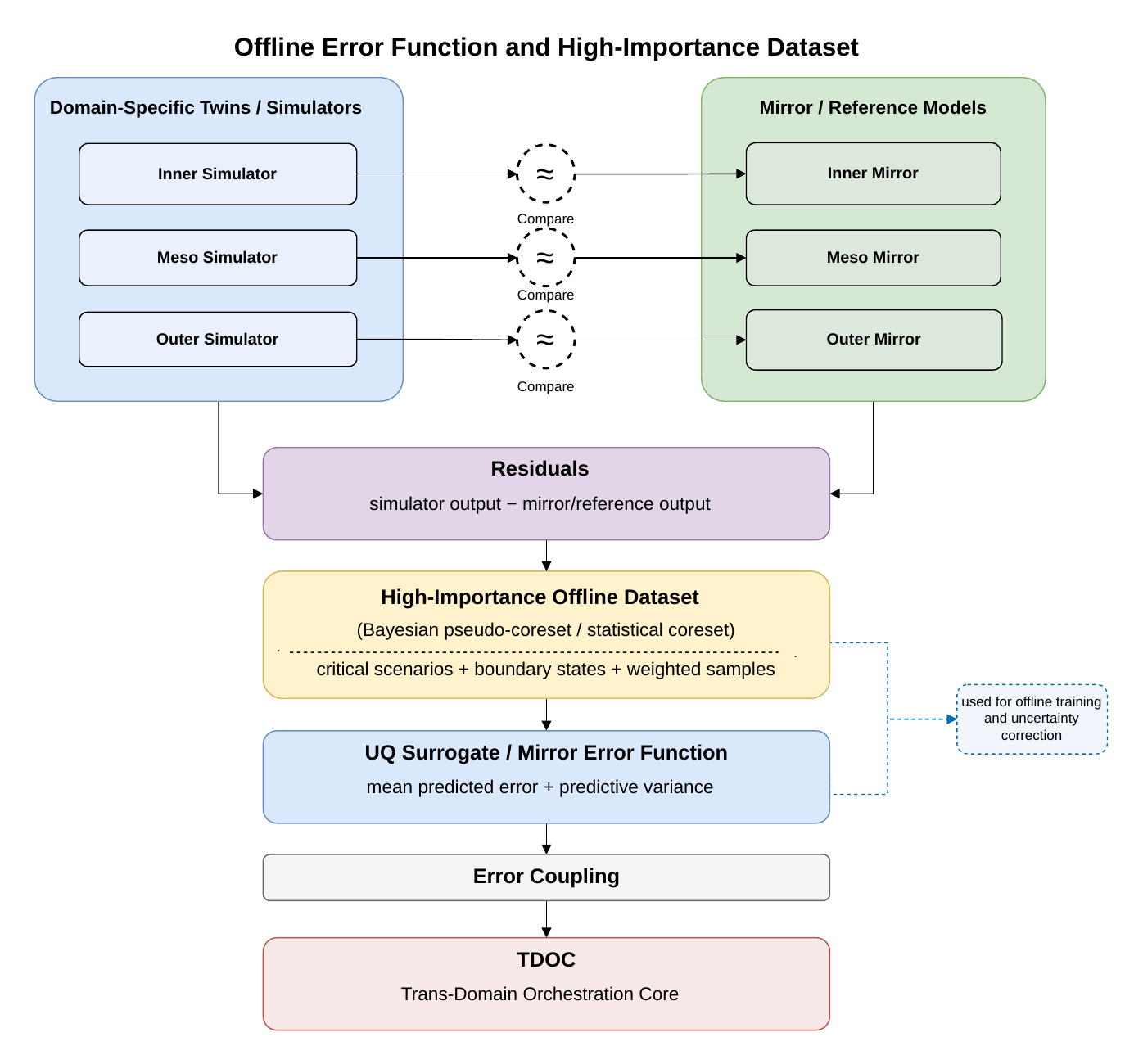}
\caption{Offline Error Function and High-Importance Dataset: The outputs of the inner, meso, and outer simulators are compared with the corresponding mirror models, and the resulting residual, together with critical scenarios and boundary states, is stored in the form of a high-importance offline dataset or Bayesian pseudo-coreset. This set is used to train the mirror error function, construct the UQ surrogate, estimate the mean error and predictive variance, and then transfer the error through the Error Coupling path to TDOC.}

\label{fig:offline-error-model}
\end{figure}

Considering the scenario-based, low-volume, and structured nature of the preprocessed climate data selected as a dataset from high-importance offline states in the target domain, a lightweight uncertainty estimator can be constructed in the offline loop of Layers 2, 3, and 4. In this framework, the output of domain-specific simulators in Layer 2 is compared with the reference output, mirror output, or the output of the fused model in Layers 3/4, and the resulting residual, together with critical scenarios and boundary states, is transferred to the high-importance offline dataset or Bayesian pseudo-coreset; this set is then used as training data for the mirror error function and the Uncertainty Quantification (UQ) surrogate (Figure \ref{fig:offline-error-model}). The combination of Uncertainty Quantification methods with process approximations creates a fast model surrogate that estimates the predicted mean error and predictive variance for each new scenario.

In this case, predictive variance is an indicator of increased uncertainty, departure from the validity domain, or the probability of error transfer among domains; therefore, Coupling Error detection is performed through the combination of residual, predictive variance, and the dependency pattern among the outputs of domain-specific twins \cite{escriba2024digital}. Models such as sparse Gaussian process (GP) / inducing points and Random Fourier Features (RFF) are candidates for these process approximations \cite{mohammadi2024emulating, titsias2009variational}.

The high-importance offline dataset, in the role of a Bayesian pseudocoreset or statistical coreset, is a small, weighted, and representative subset of the state space that summarizes the posterior distribution of system parameters with minimal approximation error relative to the full data \cite{manousakas2020bayesian}. These data can be selected through methods such as one-shot sampling, active learning, global coverage of the space, and information maximization in sensitive regions \cite{polke2026adaptive}. In practice, this set acts as a compact database of residuals, outputs of high-fidelity simulators, and critical/boundary scenarios, and, with the help of a surrogate model, enables fast estimation of uncertainty and error transfer through the Error Coupling path to TDOC without rerunning the full simulator \cite{mohammadi2024emulating, semenova2025case}.

\subsection{Error Feedback and Model Correction at Runtime}

In the online phase, runtime error is generated after the decision is applied in the real execution environment from Layer 6 and is transferred to Layer 7. This feedback from the physical system to the virtual model is considered necessary for validation, updating, completing the simulation model, and increasing prediction accuracy \cite{jones2020characterising, thelen2022comprehensivepart1, birk2022automatic}. This feedback can correct the shared state, the parameters of domain-specific twins, the confidence level, and uncertainty, and, in the proposed TDDT framework, in an extended form, it also updates the weights of the objective function, decision constraints, and the coupling intensity among domains \cite{thelen2022comprehensivepart1, thelen2022comprehensivepart2, birk2022automatic}.

In general, the online trans-domain error after decision execution in the real system is defined as follows:

\begin{equation}
\label{eq:online_td_error}
e_{TD}^{online}(t) = y_{real}(t) - y_{TDDT}(t)
\end{equation}

The error in Equation~\eqref{eq:online_td_error} is traced by the condition ($\mathrm{Route}(e_{TD})$) in Equation~\eqref{eq:error_routing}; that is, the feedback return path is differentiated between Layer 2 and Layer 3 depending on the dominant location where the error is generated. If the error originates from the model, parameter, simulator, or state of a specific twin, the feedback returns to Layer 2; however, if the error is generated after state alignment, construction of the shared state, semantic/time synchronization, coupling, constraints, objectives, or shared uncertainty, the feedback is transferred to Layer 3. In the case of a combined error, simultaneous or sequential correction of Layers 2 and 3 is performed.

\begin{equation}
\label{eq:error_routing}
\mathrm{Route}(e_{\mathrm{TD}}) =
\begin{cases}
\mathrm{Layer\ 2}, &
\text{if } e_{\mathrm{domain}} > \tau_{\mathrm{domain}}, \\[4pt]

\mathrm{Layer\ 3}, &
\text{if } e_{\mathrm{alignment}} + e_{\mathrm{coupling}} + e_{\mathrm{sync}}
> \tau_{\mathrm{TD}}, \\[4pt]

\mathrm{Layer\ 2+3}, &
\text{if both error groups exceed their thresholds.}
\end{cases}
\end{equation}

Therefore, model correction at runtime can be represented as follows (Equation~\eqref{eq:td_parameter_update}):

\begin{equation}
\label{eq:td_parameter_update}
\theta_{\mathrm{TD}}(t+1)
=
\theta_{\mathrm{TD}}(t)
+
\Delta \theta
\left(
e_{\mathrm{TD}}^{\mathrm{online}}(t),
S_{\mathrm{TD}}(t),
U_{\mathrm{TD}}(t),
\mathrm{Route}(e_{\mathrm{TD}})
\right)
\end{equation}

In this relation, ($\theta_{TD}$) denotes the parameters of models, weights, constraints, and couplings; ($e_{TD}^{online}$) is the observed error after real execution; ($S_{TD}$) is the trans-domain shared state; ($U_{TD}$) is the system uncertainty; and ($\mathrm{Route}(e_{TD})$) is the condition that determines the feedback path between Layer 2 and Layer 3. Therefore, TDDT creates an online closed loop in which the real environment feeds the domain-specific twins; the twins evaluate the future and scenarios; the trans-domain objective function selects the decision; the decision is executed in the real system; and the returned error, depending on its origin, corrects either the domain-specific models or the fused trans-domain model for subsequent decisions \cite{jones2020characterising, thelen2022comprehensivepart1, thelen2022comprehensivepart2, birk2022automatic}.

\section{Conceptual and Architectural Distinction between CDDT and TDDT}
\label{sec:cddt-comparison}

After examining the details of the conceptual design of TDDT, we can conduct a comparison between the two approaches of TDDT and CDDT.

\subsection{Conceptual Distinction between CDDT and TDDT}

\begin{itemize}

  \item CDDT: This approach seeks to identify commonalities, differences, and reusable development processes across different domains in order to create reference models, shared development methods, and generalizable platforms for digital twins \cite{heindl2022structured, dalibor2022cross}. In this case, domains maintain their independent identity, and integration is mainly performed at the levels of data, semantics, model architecture, and software structure. Its key objective is to answer the question of how digital twins in different fields can be made comparable, reusable, interoperable, and, as far as possible, standardized.
  
  \item TDDT, with a focus on operational integration and a unified dynamic system: This concept, within the proposed framework of this article, is close to the literature on composite complex systems or System-of-Systems \cite{mour2013agent, maier1998architecting, delaurentis2005understanding} and goes beyond the comparison or semantic connection of domains; its objective is to construct a unified operational system from heterogeneous domains \cite{amiri2025jumeaux}. In this approach, domains are integrated into a shared causal and decision-making dynamic loop, such that the output, error, or behavior of one domain can directly affect the control, prediction, and dynamics of another domain and lead, at the whole-system level, to emergent behaviors and simultaneous multi-domain decision-making.

\end{itemize}

In fact, CDDT asks: “How can digital twins in different domains be compared, standardized, made reusable, and rendered interoperable with one another?”

Whereas TDDT asks: “How can several heterogeneous domains be connected within a single digital twin in such a way that the output, error, and decision of each domain change the behavior and control of other domains?”

\subsection{Layered Architectural Distinction between CDDT and TDDT}

This section shows their difference in terms of layered arrangement and the flow path of data, model, state, decision, and control. In a CDDT, the layers usually include physical/domain systems, independent twins of each domain, the interoperability and semantic mapping layer, the reference model or reusable components layer, and the comparative applications or decision-support layer. The objective of this architecture is to discover common patterns, standardize, share data, develop general-purpose tools, and increase reusability across domains \cite{dalibor2022cross, ferko2022architecting}.

\begin{figure}[!htbp]
\centering
\includegraphics[width=0.75015\linewidth,alt={Side-by-side layered comparison of cross-domain interoperability and trans-domain operational coupling with feedback.}]{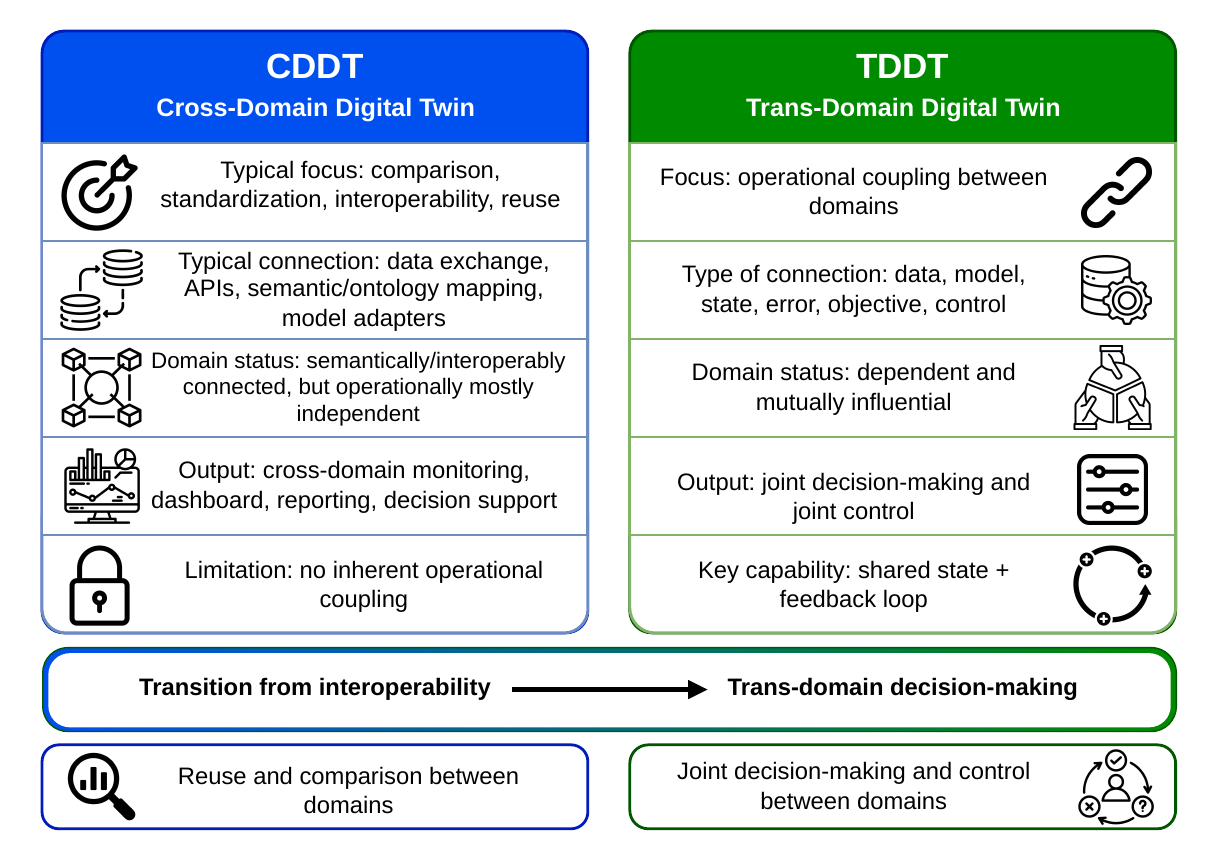}
\caption{Layered Distinction between CDDT and TDDT: On the CDDT side, domain-specific twins are mainly connected through the layers of interoperability, semantic mapping, reference models, and comparative/decision-support applications; however, on the TDDT side, the outputs of domain-specific twins enter the fused trans-domain model and, through the shared state, operational coupling, temporal synchronization, decision-making, and online feedback, lead to multi-domain control or correction.}

\label{fig:cddt-tddt-comparison}
\end{figure}

In contrast, in TDDT, the intermediate layer is transformed into a fused trans-domain model that creates the shared state, operational coupling, and temporal synchronization; therefore, the output of one domain is not used only for comparison or interoperability, but can change the state, error, constraint, objective, decision, or control of another domain \cite{devrieze2024federated}. Accordingly, Figure \ref{fig:cddt-tddt-comparison} should be read as the architectural representation of this transition: the transition from CDDT layers based on interoperability and reuse to TDDT layers based on fused state, operational coupling, feedback, and trans-domain control.

This distinction is also consistent with the literature on composite and federated DTs; for example, in federated simulation, a production line can be constructed from several independent DTs with separate simulation models, but this composition still requires a formal mechanism for model interaction and composition \cite{devrieze2024federated}.

\section{Conceptual Validation of the Proposed TDDT Framework}
\label{sec:validation}

The validation presented in this section is conceptual and based on internal consistency, requirements traceability, and scenario-based walkthrough; therefore, it does not replace verification, validation, uncertainty quantification, or field evaluation of a real implementation \cite{jones2020characterising, shao2023credibility, sel2025survey}. The objective is to examine whether the architectural components can coherently address the gaps and requirements defined in Sections~\ref{sec:definition} and~\ref{sec:architecture}.

\subsection{Scope and Level of Validation}

Validation is conducted at three levels:

\begin{enumerate}

  \item \textbf{Structural} : the presence of the required architectural components;
  
  \item \textbf{Behavioral} : the capability to transfer effects among domains;

  \item \textbf{Decision-related} : the use of multiple domains in decision-making and feedback.

\end{enumerate}

The numerical validity of the models, prediction accuracy, and operational safety are outside the scope of this conceptual evaluation and must be examined in future implementations \cite{shao2023credibility, sel2025survey, deantoni2024quantifying}.

\subsection{Structural Validation and Requirements Traceability}

\begin{figure}[!htbp]
\centering
\includegraphics[width=0.91846\linewidth,alt={Traceability matrix linking minimum architectural requirements to the components responsible for satisfying them.}]{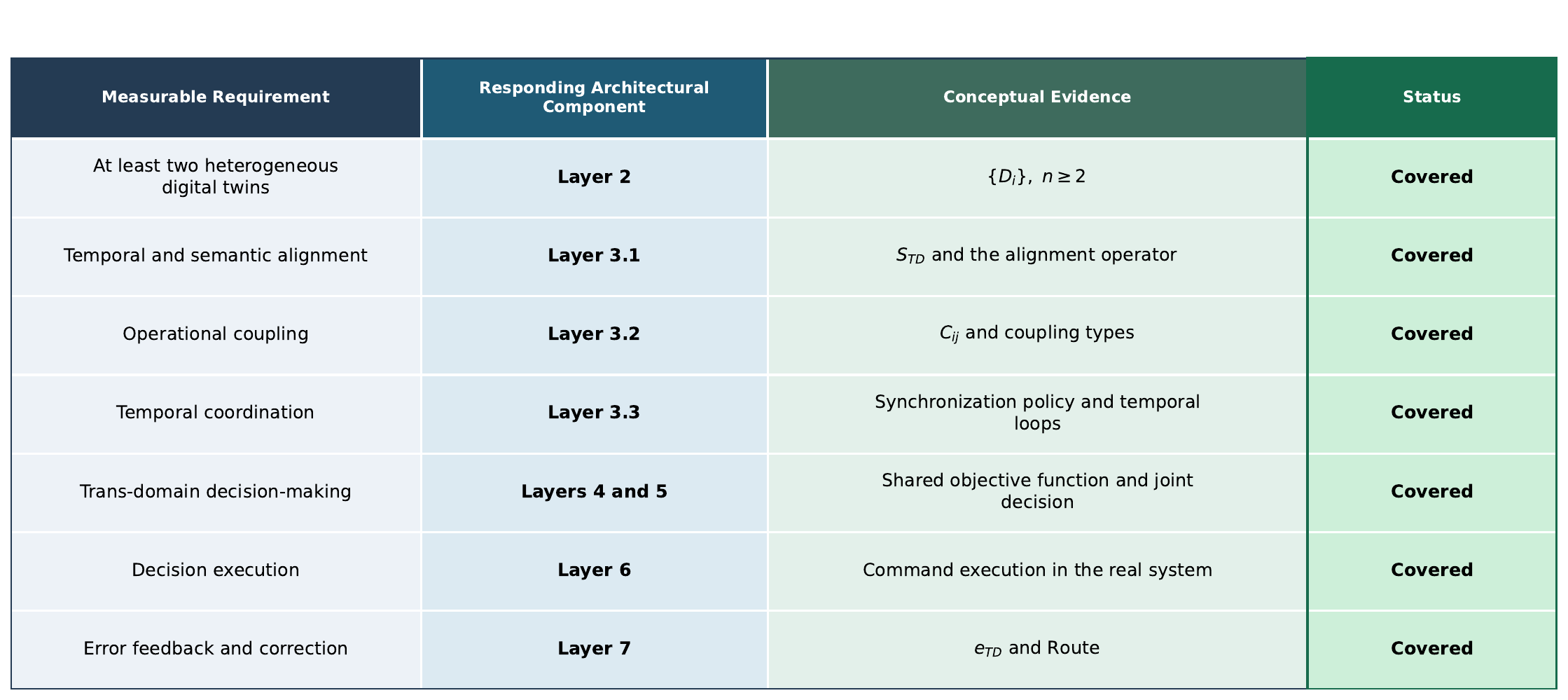}
\caption{Conceptual traceability matrix between the minimum TDDT requirements and the corresponding architectural components. The matrix shows how heterogeneous twins, state alignment, operational coupling, temporal coordination, trans-domain decision-making, execution, and error feedback are represented in the proposed architecture. STD in the figure denotes the shared trans-domain state ($S_{\mathrm{TD}}$).}

\label{fig:requirements-traceability}
\end{figure}

To examine internal consistency, each research gap and minimum requirement is mapped to its corresponding architectural component. Figure \ref{fig:requirements-traceability} presents a conceptual mapping between the minimum TDDT requirements and the architectural components proposed in this article.

This mapping only demonstrates the conceptual completeness of the architecture and does not prove the correctness, adequacy, or practical performance of each component.

The framework is considered conceptually consistent if:
\begin{enumerate}
    \item each minimum requirement is addressed by at least one architectural component;
    
    \item no trans-domain decision is generated without data or states from multiple domains;
    
    \item each coupling has a specified source, destination, type, and temporal scale;
    
    \item each observed error has a defined correction pathway.
\end{enumerate}

\subsection{Walkthrough of a Representative Scenario}

In an enclosed livestock facility, the climate, biology/growth, energy, and control twins are considered heterogeneous domains. An increase in temperature and gas concentration changes the climate state; growth status and feed intake update the biological constraints; and energy cost affects the decision function. These outputs are aligned in ($S_{TD}$) and enter the ventilation decision through state, objective, and control coupling. The actual system response is then returned to the feedback layer as an error \cite{taheri2022model, vanderlinden2019ligaps, tabase2023ammonia}.

\begin{figure}[!htbp]
\centering
\includegraphics[width=0.92323\linewidth,alt={Livestock scenario tracing climate and gas changes through biological and energy effects, decision, execution and error feedback.}]{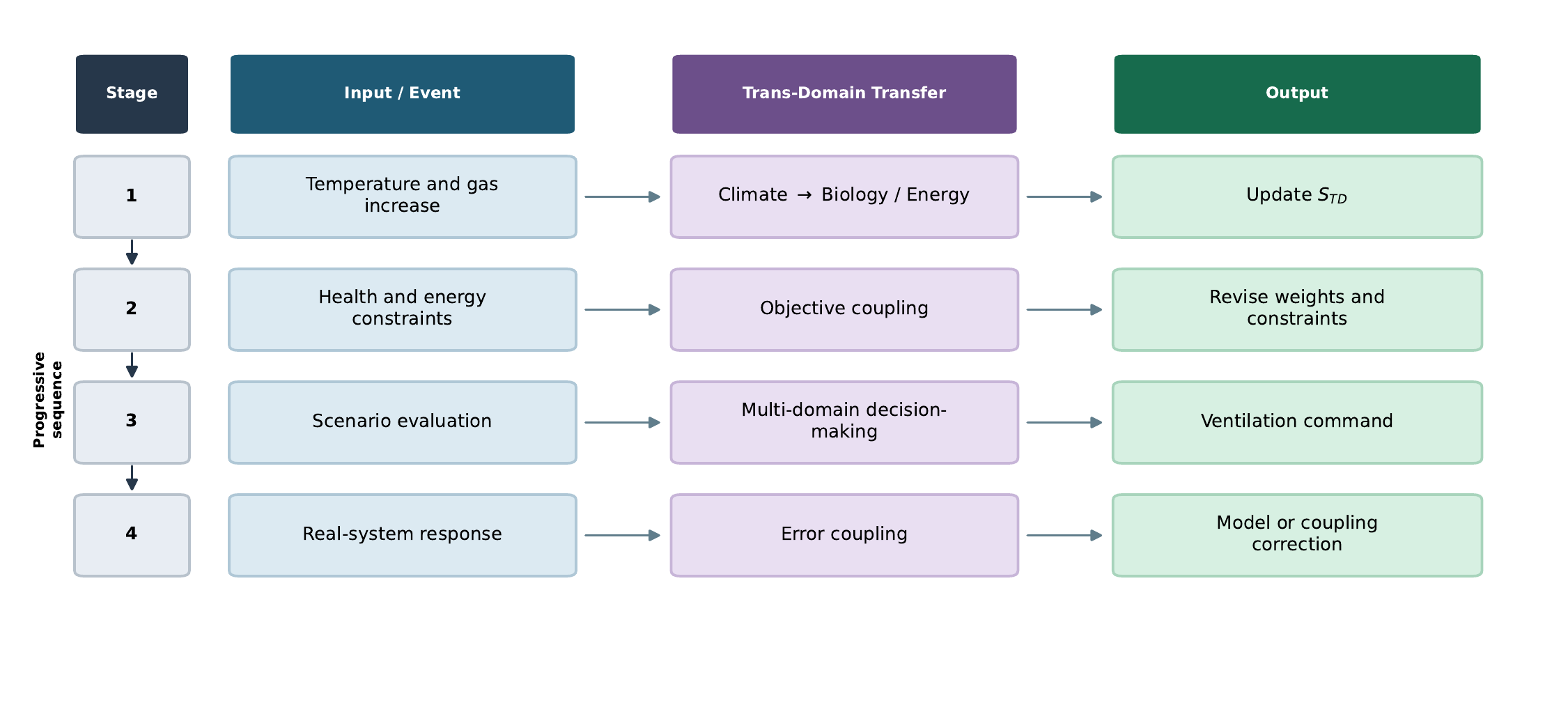}
\caption{Illustrative trans-domain walkthrough for a closed livestock environment. Temperature and gas changes propagate to the biological and energy domains, revise objectives and constraints, generate a ventilation decision, and return the observed system response through error coupling for model or coupling correction. STD in the figure denotes the shared trans-domain state ($S_{\mathrm{TD}}$).}
\label{fig:scenario-walkthrough}
\end{figure}

Figure \ref{fig:scenario-walkthrough} illustrates the proposed trans-domain sequence for the closed-livestock example, from environmental change to decision execution and error feedback.

Accordingly, four tests are proposed:

\begin{itemize}
    \item \textbf{Removal of the shared state} : decisions are restricted to unaligned information;
    \item \textbf{Removal of objective/control coupling} : the system is reduced to multiple parallel twins;
    \item \textbf{Removal of temporal coordination} : domain outputs may become inconsistent;
    \item \textbf{Removal of feedback} : errors are not corrected in subsequent cycles.
\end{itemize}

Failure in any test indicates that the removed component has a structural role in the proposed TDDT formulation; however, the magnitude of its effect must be determined through ablation or practical experimentation.

\subsection{Acceptance Criteria and Validation Limitations}

\begin{figure}[!htbp]
\centering
\includegraphics[width=0.78492\linewidth,alt={Conceptual acceptance matrix covering completeness, traceability, coordination, joint decisions, feedback and uncertainty.}]{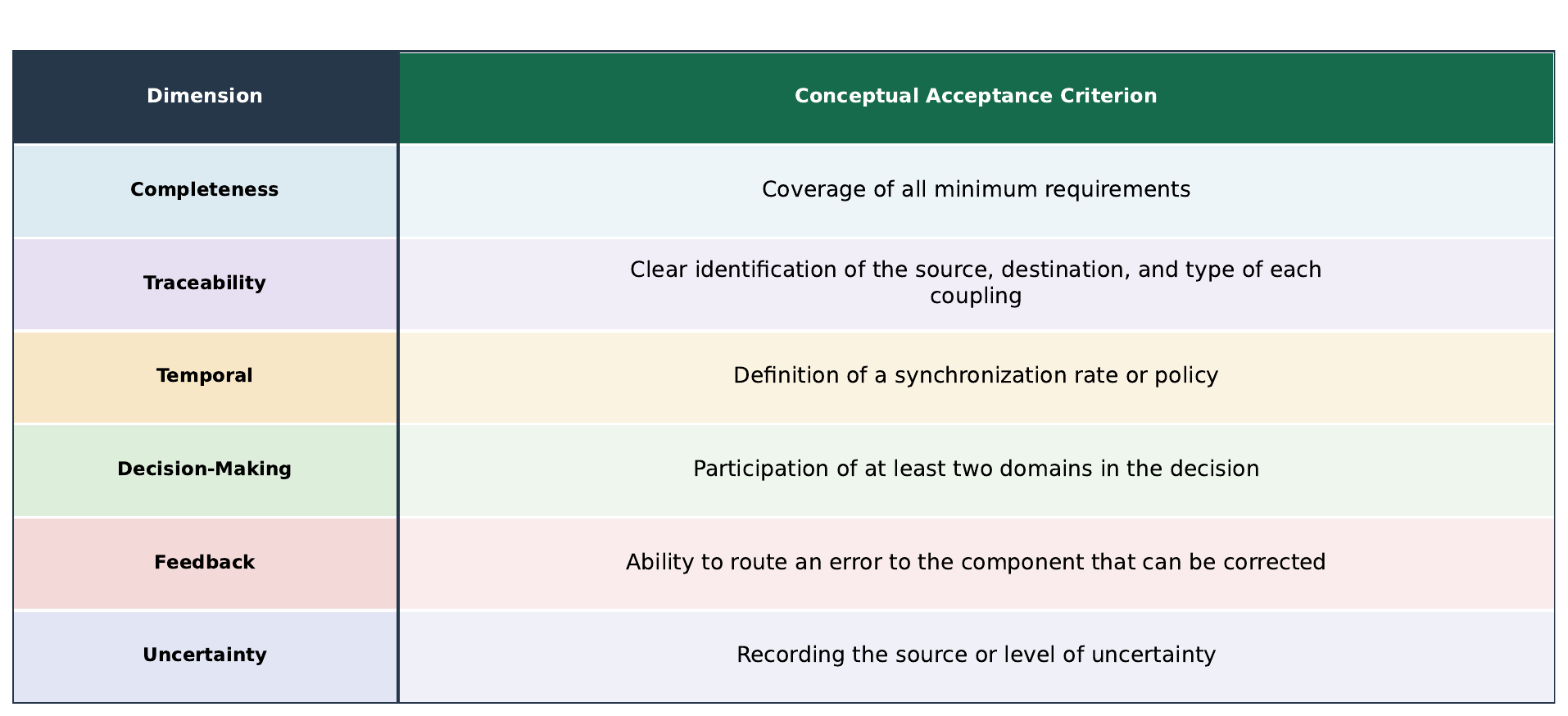}
\caption{Proposed conceptual acceptance dimensions for TDDT. The criteria assess completeness, coupling traceability, temporal coordination, multi-domain participation in decision-making, feedback routing, and uncertainty recording; satisfying them indicates conceptual conformity rather than practical validity.}

\label{fig:acceptance-criteria}
\end{figure}

Figure \ref{fig:acceptance-criteria} summarizes the proposed conceptual acceptance dimensions used to assess the internal conformity of a TDDT architecture.

Satisfaction of these criteria only indicates conceptual compliance with the proposed architecture. Practical validity requires code and model verification, validation using real-world data, uncertainty quantification, sensitivity analysis, robustness testing, and safety evaluation \cite{shao2023credibility, sel2025survey,deantoni2024quantifying, nist2024digital}.

\section{TDDT Maturity Model}
\label{sec:maturity}

The maturity model presented in this section is a proposed classification for describing the transition from independent twins to adaptive TDDT. A higher level is not necessary or desirable for every application, and the target level should be selected based on risk, cost, data, and decision-making requirements \cite{kritzinger2018digital,jones2020characterising,thelen2022comprehensivepart1,verdouw2021digital,michael2022integration,fuller2020digital,khedr2025composition,onaji2022digital,tao2018digital}. This model differs from the Monitoring–Predictive–Prescriptive–Autonomous functional classification; a DT may be functionally autonomous, yet without coupling among domains, it is not considered a TDDT \cite{kritzinger2018digital,verdouw2021digital,fuller2020digital,wuni2026conceptualising}.

\begin{figure}[!htbp]
\centering
\includegraphics[width=0.94396\linewidth,alt={Maturity scale from L0 independent twins to L6 bounded trustworthy adaptation, with minimum conformity at L3.}]{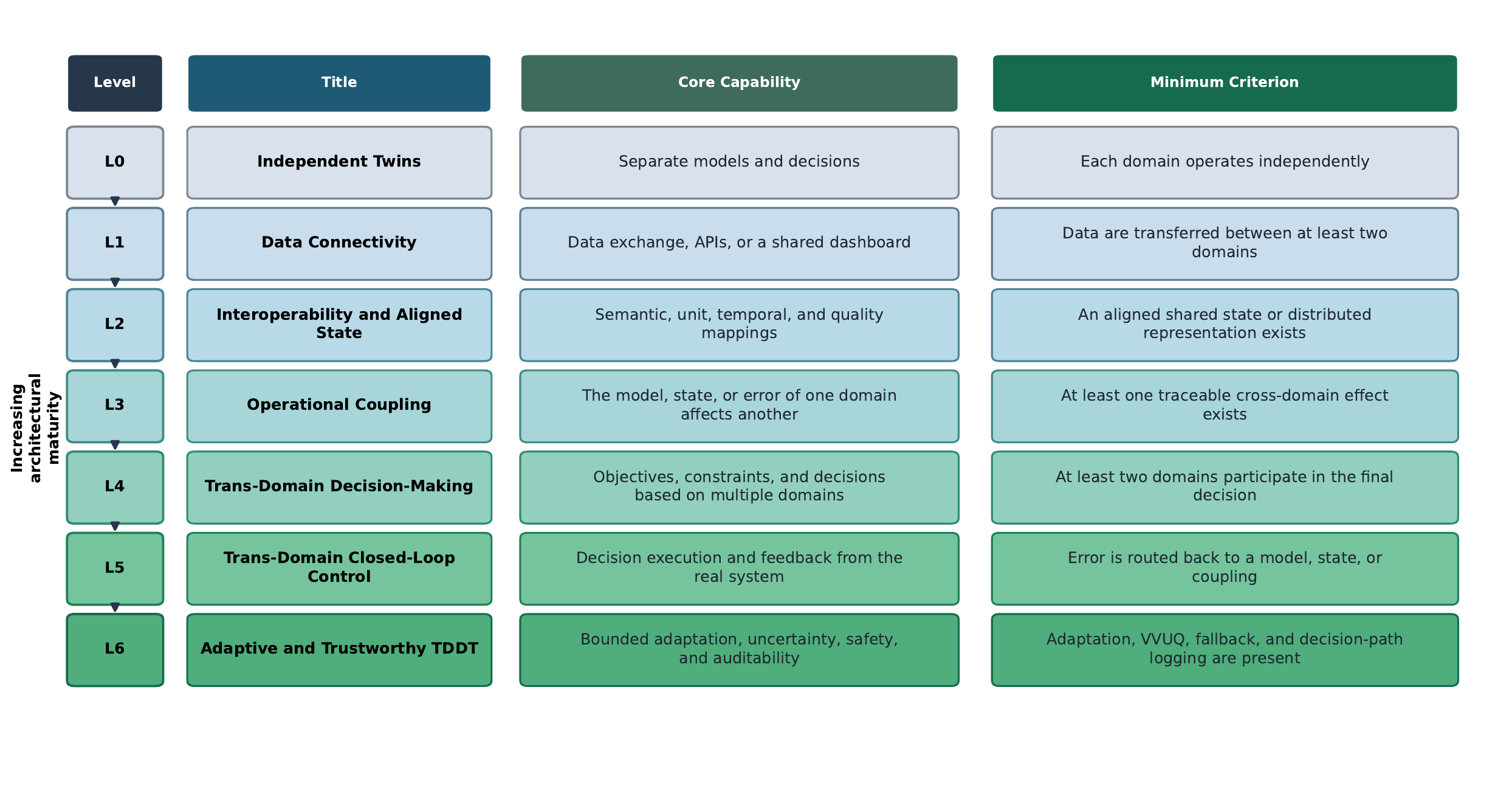}
\caption{Proposed maturity levels for TDDT implementations. Levels L0–L2 represent independent, connected, and interoperable multi-domain prerequisites; minimum conformity with the proposed TDDT definition begins at L3 with traceable operational coupling, while L4–L6 progressively add joint decision-making, closed-loop feedback, and bounded trustworthy adaptation. VVUQ denotes verification, validation, and uncertainty quantification.}

\label{fig:maturity-model}
\end{figure}

Figure \ref{fig:maturity-model} presents the proposed architectural maturity levels from independent domain twins to adaptive and trustworthy TDDT.

Levels L0 to L2 are multi-domain prerequisites, and minimum compliance with TDDT begins at L3. Advancement should be based on evidence of alignment, coupling validity, joint decision-making, feedback, and assurance; the mere presence of an API, dashboard, automated control, or the TDOC designation is not sufficient for level advancement. Moreover, a higher level does not necessarily guarantee greater accuracy, safety, or practical value and must be assessed using the evaluation protocol of the article \cite{michael2022integration,khedr2025composition,oleynikova2017voxblox}.

\section{TDDT Deployment Architecture and Software Components}
\label{sec:deployment}

The architecture presented in this section is a proposed reference view for mapping the conceptual layers of TDDT to executable components. The components may be deployed locally, in an edge–cloud configuration, centrally, or in a distributed manner, provided that interoperability, temporal coordination, coupling, decision-making, and feedback are preserved \cite{michael2022integration, david2025interoperability, vergara2024federated, khedr2025composition, budiardjo2021digital}.

The logical view includes a device and safety gateway, a message/event bus, domain-twin services, a model adapter, TDOC or an equivalent orchestrator, a shared-state and time service, an ontology service, prediction and decision services, a time-series store, a model registry, a knowledge store, and monitoring/audit. It is not necessary for all components to exist as independent services, and lightweight implementations may integrate multiple roles.

\begin{figure}[!htbp]
\centering
\includegraphics[width=0.44369\linewidth,alt={Reference software deployment linking gateways, message bus, twin services, orchestration, prediction, decisions and supporting stores.}]{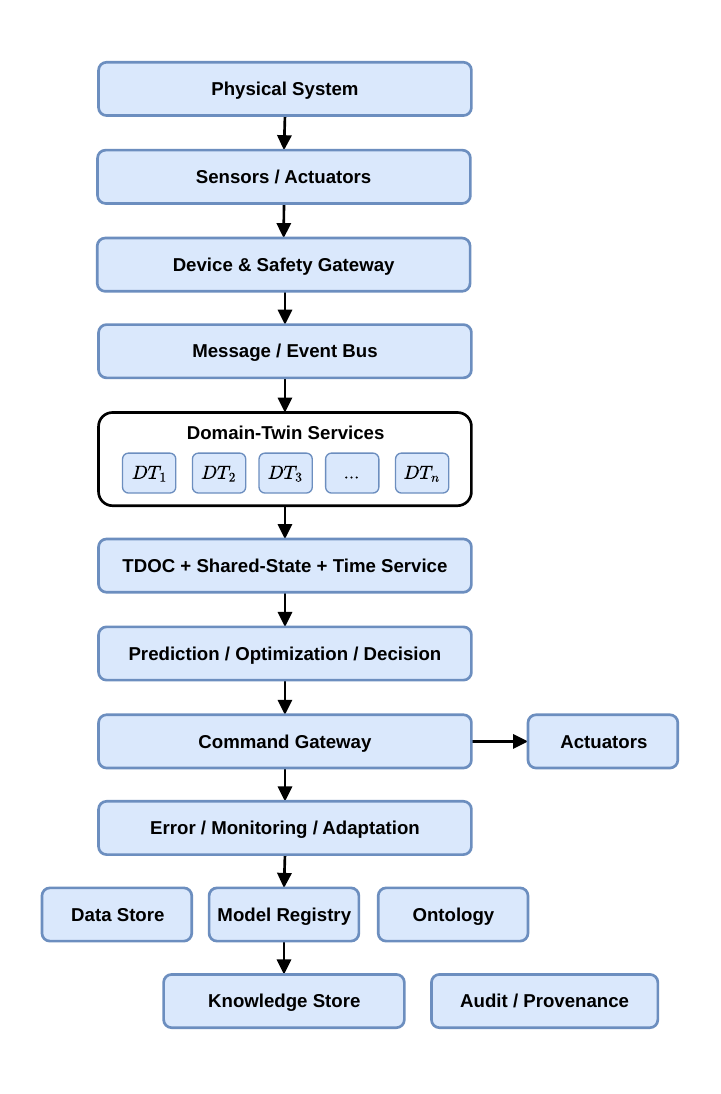}
\caption{Proposed reference deployment view of the TDDT architecture. Sensor and actuator data pass through the device and safety gateway and message bus to the domain-twin services; TDOC coordinates the shared state, timing, prediction, and decision services, while monitoring, adaptation, data storage, model registration, ontology, knowledge storage, and provenance support the operational lifecycle. Components may be deployed centrally, at the edge, or in a distributed configuration.}

\label{fig:deployment-architecture}
\end{figure}

Figure \ref{fig:deployment-architecture} shows a reference deployment view that maps these logical roles to the main runtime and supporting software components.

In the online flow, sensor data are transferred to the domain twins after timestamping, unit conversion, and quality control; TDOC aligns and couples the outputs and sends the decision to the safety gateway; the actual response is then returned to correct the model, state, or coupling. TDOC is a logical role and may be implemented centrally or in a distributed manner.

FMI/FMU can be used for model packaging, HLA for distributed simulation, and ISO 23247 for the conceptual mapping of roles \cite{modelica2024fmi302, ieee2025hla1516, iso2021iso23247}. Data contracts should retain at least the identifier, timestamp, unit, quality, uncertainty, version, and context \cite{david2025interoperability, budiardjo2021digital, correia2023data, karabulut2023ontologies}. This mapping does not imply formal compliance.

Low-latency and safety-critical tasks are preferably executed close to the edge, while computationally intensive training, archiving, and model management may be placed on the central platform. Each decision should be traceable to the active versions of the model, data, coupling, and parameters. Latency, availability, scalability, security, observability, and recoverability should also be evaluated in accordance with the level of risk \cite{correia2023data, shao2023credibility, elhajj2024systematic, kustelega2024privacy}.

\section{TDDT Evaluation Criteria and Protocol}
\label{sec:evaluation}

The criteria presented in this section do not constitute a formal benchmark for all TDDTs, but rather the article’s proposed minimum protocol for evaluating future implementations. The type of metric, baseline, and acceptance threshold should be determined according to the application, risk level, data, and degree of automation \cite{jones2020characterising, thelen2022comprehensivepart2, shao2023credibility, sel2025survey}. The evaluation should separately examine the domain models, shared state, coupling, temporal coordination, decision-making, uncertainty, robustness, computational cost, and end-to-end performance, because validity at one level does not guarantee validity at the other levels \cite{shao2023credibility, sel2025survey, deantoni2024quantifying}.

\subsection{Evaluation Dimensions and Criteria}

\begin{figure}[!htbp]
\centering
\includegraphics[width=0.88000\linewidth,alt={Multidimensional evaluation matrix spanning model validity, alignment, coupling, timing, prediction, control, uncertainty and performance.}]{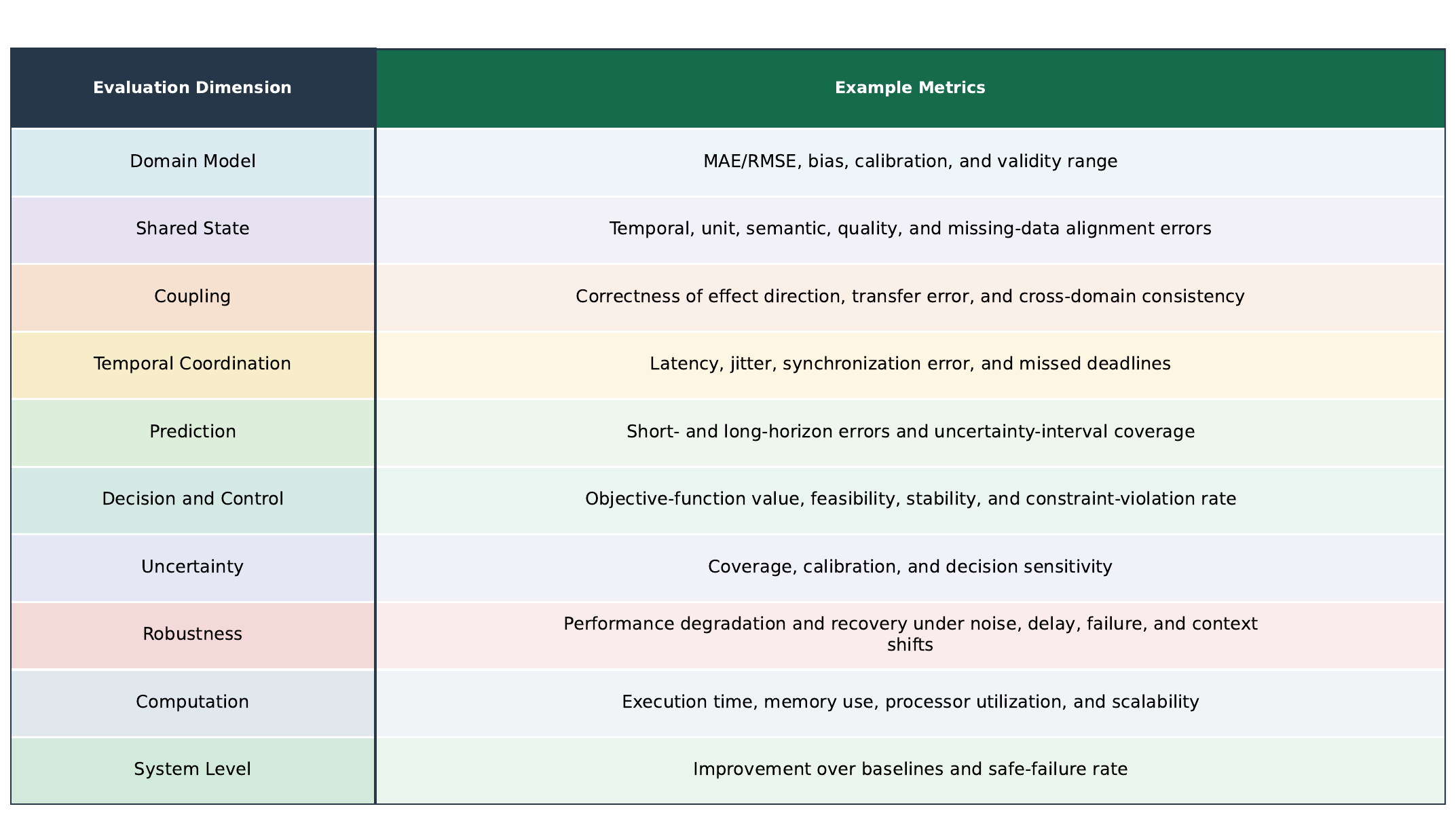}
\caption{Proposed evaluation dimensions and example metrics for TDDT implementations. The dimensions separately address domain-model validity, shared-state alignment, coupling, temporal coordination, prediction, decision and control, uncertainty, robustness, computational performance, and system-level improvement. The listed metrics are illustrative and should be selected according to the application and risk level. MAE denotes mean absolute error, and RMSE denotes root mean square error.}
\label{fig:evaluation-metrics}
\end{figure}

Figure \ref{fig:evaluation-metrics} summarizes the proposed evaluation dimensions and representative metrics for future TDDT implementations.

These metrics are illustrative, and not all of them are mandatory in every application; however, the omission of any dimension should be justified \cite{thelen2022comprehensivepart2, shao2023credibility, sel2025survey}.

The following relations can be used to evaluate certain TDDT-specific components:

\begin{equation}
\label{eq:constraint_violation_ratio}
R_{\mathrm{cv}}
=
\frac{
N_{\mathrm{violated}}
}{
N_{\mathrm{evaluated}}
}
\end{equation}

which (Equation~\eqref{eq:constraint_violation_ratio}) represents the constraint violation rate, and:

\begin{equation}
\label{eq:td_gain}
G_{\mathrm{TD}}
=
\frac{
M_{\mathrm{TDDT}}
}{
M_{\mathrm{baseline}}
}
\end{equation}

which (Equation~\eqref{eq:td_gain}) represents the gain or change in TDDT performance relative to the baseline. The direction of the difference should be determined according to the nature of the metric, such as benefit, cost, or error.

\subsection{Minimum Evaluation Protocol}

The proposed protocol includes the following steps:

\begin{enumerate}
    \item Define the application domain, objectives, constraints, and risk level;
    \item Determine the reference data, scenarios, and validity range;
    \item Independently validate each domain twin;
    \item Evaluate the shared state and data alignment;
    \item Separately test each coupling;
    \item Evaluate temporal coordination and latency;
    \item Evaluate decision-making, feasibility, and constraint satisfaction;
    \item Compare against baselines;
    \item Conduct ablation, sensitivity analysis, and stress testing;
    \item Report uncertainty, computational resources, limitations, and reproducibility.
    
\end{enumerate}

The acceptance criterion for each stage should be defined before testing and based on the application risk, and should not rely solely on a general threshold \cite{shao2023credibility, sel2025survey}.

\subsection{Baselines and Ablation Testing}

The minimum baselines should include independent domain twins, a multi-domain architecture with data exchange but without operational coupling, TDDT without online adaptation, and the conventional application-specific decision-maker. The superiority of TDDT should be claimed only when an improvement in the target metric is observed without an unjustified increase in constraint violations, uncertainty, risk, or computational cost.

To determine the contribution of the architectural components, the complete system should, where possible, be compared with versions without a shared state, without state coupling, without objective/control coupling, without online feedback, and without an uncertainty model. Removal of the meso loop or single-episode training mechanisms should be examined only in implementations that include these components, because these components are not general requirements for all TDDTs.

\subsection{Robustness and Uncertainty Testing}

Robustness should be tested at least against sensor noise, delay, missing data, model error, coupling error, failure of a twin, and context changes. In each test, performance degradation, decision changes, error propagation, recovery time, and fallback behavior should be reported \cite{shao2023credibility, sel2025survey, deantoni2024quantifying, nist2024digital}.

Uncertainty should also be distinguished at least by its data, model, coupling, and decision sources. Its evaluation should include coverage, calibration, sensitivity analysis, and examination of decision changes under uncertainty; a narrow interval alone is not an indication of validity \cite{kennedy2001bayesian, thelen2022comprehensivepart2, sel2025survey, deantoni2024quantifying}.

\subsection{Temporal Performance and Results Reporting}

In online applications, in addition to the mean execution time, high latency percentiles, jitter, missed deadlines, memory consumption, and changes in cost as the number of twins increases should be reported. The use of surrogate models, Sparse GP, or computational-cost reduction methods should be evaluated together with the potential loss of accuracy and changes in uncertainty \cite{mohammadi2024emulating, titsias2009variational, manousakas2020bayesian, polke2026adaptive}.

Each evaluation should report the data and scenarios, model versions, active couplings, temporal rates, baselines, metrics, uncertainty intervals, computational resources, failure tests, and limitations of generalization. Simulation results only indicate performance within the simulated scenario and should not be generalized to laboratory, field, or operational validity without independent evidence \cite{jones2020characterising, shao2023credibility, sel2025survey}.

\section{Failure Management, Fallback, and Operational Safety}
\label{sec:safety}

This section presents the proposed reference requirements for limiting the effects of failures and uncertain decisions in TDDT. The type of response and level of redundancy should be determined according to the application and the consequences of the hazard \cite{fuller2020digital, rasheed2020digital, shao2023credibility, sel2025survey}. Failures may originate from sensors, data, models, coupling, timing, networks, decisions, or adaptation, and may propagate to other domains through the shared state \cite{jones2020characterising, shao2023credibility, sel2025survey, deantoni2024quantifying}.

Before execution, the safety gateway should verify the command range, rate of change, timing, actuator health, hard constraints, and confidence. A command is executed only if it passes the safety test; otherwise, the system should revert to the last valid bounded state, exclusion of the uncertain domain, a local controller, a conservative policy, human override, or safe shutdown. The fallback mechanism should be simple, independent, and pre-validated, and its objective should be to preserve safety or minimum service rather than to continue optimal operation \cite{camacho2007model}.

In the event of a failure in a twin or coupling, its effects should be mitigated or quarantined, and the system should operate in a degraded mode. Online updates should also be bounded, versioned, and rollback-capable, and should be suspended when data are insufficient, uncertainty is high, or a constraint is violated \cite{bolender2021self, kennedy2001bayesian, thelen2022comprehensivepart2, sel2025survey, deantoni2024quantifying}. Return to normal operation should be permitted only after a health check, alignment, consistency verification, and a period of stable operation.

All failure, veto, fallback, rollback, and recovery events should be recorded together with the versions of the models, data, and active couplings. The presence of these mechanisms does not demonstrate safety, and their adequacy should be examined through hazard analysis, fault-injection testing, stress testing, VVUQ, and practical evaluation \cite{shao2023credibility, sel2025survey, deantoni2024quantifying, nist2024digital, elhajj2024systematic, kustelega2024privacy}.

\begin{figure}[!htbp]
\centering
\includegraphics[width=0.77000\linewidth,alt={Safety and recovery matrix pairing monitoring, validation, fault isolation, fallback, rollback and override with evidence.}]{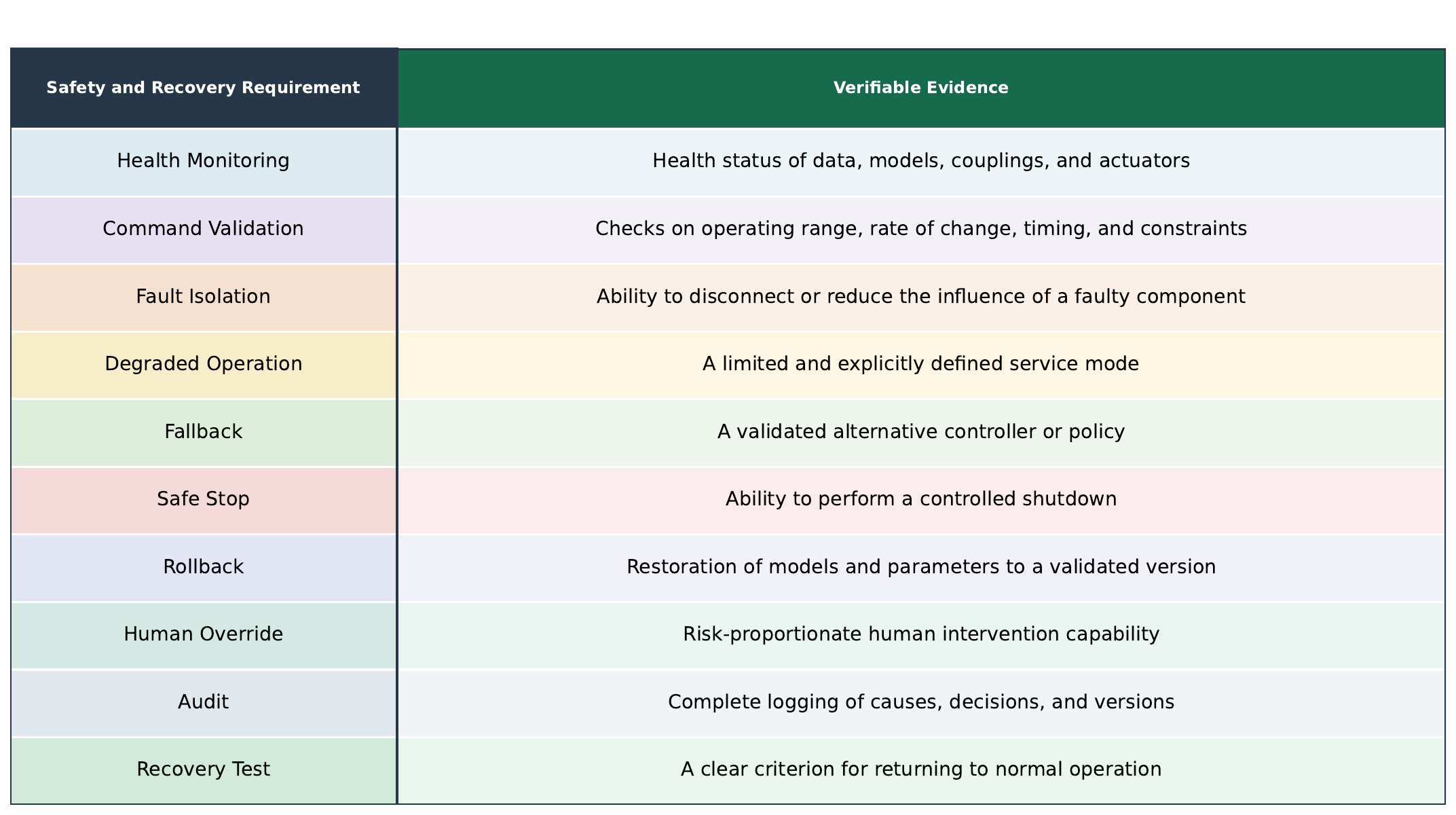}
\caption{Proposed runtime-safety, failure-management, and recovery requirements for TDDT. The table links health monitoring, command validation, fault isolation, degraded operation, fallback, safe stopping, rollback, human override, audit, and recovery testing to their expected verifiable evidence. These requirements support safety assessment but do not by themselves constitute a formal safety guarantee.}

\label{fig:safety-recovery}
\end{figure}

Figure \ref{fig:safety-recovery} summarizes the proposed runtime-safety, failure-management, fallback, and recovery requirements and their corresponding verifiable evidence.

\section{Provenance, Versioning, and Model Lifecycle}
\label{sec:provenance}

The requirements presented in this section constitute the proposed minimum mechanisms for tracing data, models, coupling, and decisions in TDDT \cite{jones2020characterising, correia2023data, shao2023credibility, iso2021iso23247}. Each important output should, as far as possible, be traceable to the data source, timestamp, model version, parameters, active couplings, uncertainty, and decision version. This traceability is necessary for decision reconstruction and rollback, but it does not guarantee the correctness of the decision \cite{correia2023data, shao2023credibility}.

Versioning should cover data, domain models, ontology, coupling, constraints, the objective function, and training knowledge. Each version should have an identifier, time, reason for change, validity range, and lifecycle status. The proposed lifecycle includes Draft, Verified, Validated, Approved, Deployed, Monitored, and Retired; the names of the stages may differ, but experimental and operational versions should be distinguished from one another \cite{shao2023credibility, sel2025survey}.

\begin{figure}[!htbp]
\centering
\includegraphics[width=0.88954\linewidth,alt={Lifecycle flow from draft through verification, validation, approval, deployment and monitoring to update or retirement.}]{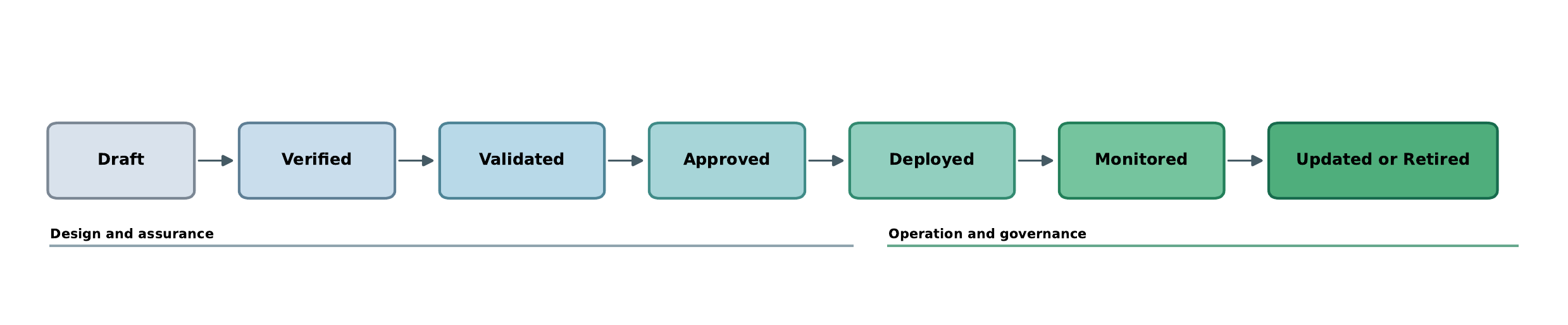}
\caption{Proposed lifecycle of a domain model or coupling in TDDT. A candidate component progresses from draft through verification, validation, approval, deployment, and monitoring, after which it may be updated or retired according to performance, uncertainty, validity-range, and governance evidence.}

\label{fig:model-lifecycle}
\end{figure}

Figure \ref{fig:model-lifecycle} illustrates the proposed lifecycle of a domain model or coupling from initial development to operational monitoring, updating, or retirement.

A new model or coupling should be released only after evaluation using independent data, uncertainty, constraints, latency, and comparison with the active version \cite{kennedy2001bayesian, thelen2022comprehensivepart2, shao2023credibility, sel2025survey}. The operational version should be monitored for drift, residuals, calibration, and context changes, and in the event of performance degradation, constraint violation, or inconsistency, mechanisms should be available for confidence reduction, shadow mode, fallback, or rollback to the latest valid version \cite{thelen2022comprehensivepart2, shao2023credibility, sel2025survey, deantoni2024quantifying}.

Each decision should be recorded together with the versions of the models, couplings, shared state, objective function, constraints, uncertainty, and the safety gateway result. The retention, access, archiving, and deletion policies for these records should comply with confidentiality and legal requirements \cite{correia2023data, elhajj2024systematic, kustelega2024privacy}. The presence of provenance and versioning alone does not demonstrate validity, safety, or accountability and should be evaluated in terms of correctness, security, and reconstructability.

\section{Threats to Validity and Scope of Claims}
\label{sec:validity}

In this article, TDDT is a proposed conceptual formulation and is not yet considered a generally established term or standard. Its boundary with Composite/Federated Digital Twin and Digital Twin System-of-Systems may overlap in some implementations; therefore, the distinction presented here is based on the criteria proposed in this article, including a shared state, operational coupling, temporal coordination, trans-domain decision-making, and traceable feedback \cite{heindl2022structured, dalibor2022cross, michael2022integration, david2025interoperability, vergara2024federated, khedr2025composition}.

Mapping the requirements to the architectural layers demonstrates conceptual consistency, but does not prove the causal validity, stability, or adequacy of the couplings. The direction and magnitude of effect and error transfer may depend on data, models, context, and delay and should be validated separately for each application \cite{thelen2022comprehensivepart2, shao2023credibility, sel2025survey, deantoni2024quantifying}.

The applications presented in the article are illustrative examples and do not constitute empirical evidence for generalizing TDDT to all domains. Moreover, the article does not include a general benchmark, a complete implementation, or an end-to-end field comparison; therefore, no superiority in terms of accuracy, safety, cost, latency, or robustness can be concluded \cite{jones2020characterising, shao2023credibility, sel2025survey, deantoni2024quantifying, nist2024digital}. The capabilities of single-episode training, CRP, SARG, MRG, Knowledge Store (KStore), hidden loop current, and online adaptation are also proposed mechanisms, and their general effectiveness requires independent evaluation.

To mitigate these threats, each twin and coupling should be validated separately, and the complete system should be evaluated using baselines, ablation, sensitivity analysis, stress testing, VVUQ, and field data. Simulation results should not be generalized to operational validity without independent evidence \cite{shao2023credibility, sel2025survey, deantoni2024quantifying, nist2024digital}.

\section{Some Potential Application Domains of TDDT as a Conceptual Framework}
\label{sec:applications}

The cases presented in this section are examples of complex systems in which multiple heterogeneous domains are operationally interdependent, and a change in one domain can alter the state, error, risk, constraint, objective, or decision of another domain. The purpose of this section is to demonstrate how TDDT can support trans-domain decision-making across different applications through a shared state, data/model/error/objective/control coupling, temporal loops, and online adaptation.

\subsection{Biology–Climate–Energy-Aware Control in Cattle Fattening}

In enclosed cattle-fattening barns, the indoor climate, animal growth, feed intake, metabolic heat production, thermal comfort, and energy consumption mutually affect one another; therefore, local climate control without considering the biological state of the animals can lead to inconsistent and costly decisions \cite{taheri2022model, vanderlinden2019ligaps, tabase2023ammonia}. Within the TDDT framework, the climate, biology/growth, energy, and control twins are placed within a shared state so that indicators such as body weight, Average Daily Gain (ADG), feed intake, heat production, and growth limitations are transformed into constraints or guidance for climate-related decisions. The fast loop evaluates ventilation, heating, inlet, and fan commands using MPC, while the slow loop feeds the cumulative effects of growth, energy, and health back into the decision objectives and constraints \cite{taheri2022model}, and the Meso loop also relies on the mutual effects of gases produced inside the barn, air quality, ventilation, energy consumption, and the feed-intake behavior of the animals \cite{tabase2023ammonia}.

\subsection{Path Planning in Autonomous Vehicles}

In autonomous vehicles, path planning must transform environmental perception, obstacles, road boundaries, vehicle dynamics, energy consumption, safety, and the prediction of traffic-agent behavior into an executable path \cite{xia2024survey, yu2021model, xu2025survey}. This problem is inherently trans-domain. The selected path simultaneously affects control constraints, stability, braking, acceleration, comfort, collision risk, and interaction with vehicles or pedestrians. In TDDT, the perception/scene, map, vehicle-dynamics, agent-prediction, energy, and control twins are coupled within a shared state so that the path, control, and risk are evaluated simultaneously, and the final decision is selected based on multi-domain consequences rather than local optimization.

\subsection{Cancer and Personalized Treatment}

In personalized cancer treatment, tumor growth, the microenvironment, angiogenesis, treatment response, toxicity, drug resistance, medical imaging, and the patient’s condition are interdependent across multiple scales and heterogeneous domains; global cancer statistics also demonstrate the importance of this issue in terms of disease burden and the need for precise therapeutic decision-making \cite{bray2024global}. Various models are applied across these domains to predict tumor growth and treatment response \cite{rockne2019mathematical, jarrett2018mathematical, hormuth2020forecasting}. In TDDT, these domains are incorporated into a shared state so that objectives such as controlling tumor growth, reducing toxicity, selecting the treatment dose/schedule, and online refinement of the patient-specific model are evaluated in a coupled manner.

\subsection{Quantum Error Correction (QEC) in Quantum Processing}

In QEC, noise, decoherence, gate errors, readout errors, code selection, decoder, connectivity topology, pulse control, and hardware resources simultaneously affect the logical error rate and computational cost \cite{breuer2016colloquium, nathan2020universal, fowler2012surface, acharya2025quantum, litinski2019game}. Models of open quantum systems, noisy channels, Lindblad/master equations, stabilizer/surface codes, and resource-estimation models each describe part of this system \cite{breuer2016colloquium, nathan2020universal, fowler2012surface, acharya2025quantum, litinski2019game}. In TDDT, the noise, hardware, control, syndrome-readout, code, decoder, compilation/scheduling, and uncertainty domains are incorporated into a shared state so that reducing the logical error rate, reducing latency, selecting the code distance, and online adjustment of control parameters are pursued simultaneously.

\subsection{Tritium Self-Sufficiency and Breeding Blanket in Nuclear Fusion}

In fusion power plants, tritium self-sufficiency depends on simultaneous coupling among plasma, neutronics, blanket, heat transfer, materials, tritium chemistry, safety, the fuel cycle, and energy economics; therefore, changes in neutron flux, temperature, material degradation, or tritium recovery can alter operational constraints and design decisions \cite{abdou2015blanket, taylor2023tritium, kasada2015system}. In TDDT, the plasma, blanket, materials, thermal, fuel, safety, and economic twins are incorporated into a shared state so that indicators such as Tritium Breeding Ratio (TBR), material degradation, thermal load, safety risk, and fuel-cycle cost are used in a coupled manner within the decision loop and online adaptation.

\subsection{Optimization of Retropropulsion, Plume, and Reentry Aerodynamics}

In reusable spacecraft, retropropulsion and atmospheric re-entry involve strong coupling among propulsion, plume, aerodynamics, aerothermodynamics, the return trajectory, Guidance, Navigation, and Control (GNC), mass/fuel, structure, and thermal health; existing models often depend on fluid-dynamics equations, turbulence models, compressible flow, and heat transfer, but computational cost and uncertainty in boundary conditions limit real-time decision-making \cite{bykerk2024retropropulsion, dresia2021multidisciplinary}. In TDDT, these domains are coupled within a shared state so that, in the fast loop, thrust, trajectory, and angle of attack are adjusted to reduce heat flux, fuel consumption, and landing dispersion, while in the slow loop, the cumulative effects of heating, structural loads, and reuse damage are fed back into the decision objectives and constraints \cite{dresia2021multidisciplinary, guadagnini2022model}.

\subsection{Stability of the Microbial Community in Clean-Energy Bioreactors}

The stability of microbial communities in clean-energy bioreactors depends on temperature, pH, organic loading, retention time, feed composition, Volatile Fatty Acids (VFA), free ammonia, mass transfer, and species competition, and small operational changes can reduce $CH_4$, $H_2$, or electrical current production \cite{harirchi2022microbiological, batstone2002anaerobic, bernard2001dynamical}. Models such as Anaerobic Digestion Model No. 1 (ADM1) and AM2\footnote{A two-step anaerobic digestion model based on acidogenesis and methanogenesis.}, chemostat/Monod, Differential-Algebraic Equations (DAE), Monod–Butler–Volmer, mass balance, biofilm, and polarization models are used to describe growth, inhibition, microbial competition, and the linkage of biokinetics with electrochemistry \cite{batstone2002anaerobic, bernard2001dynamical, dudley2020competitive, kumar2022microbial, li2023model}. In TDDT, the microbial-community, feed, solution-chemistry, gas/electrochemistry, temperature/mass-transfer, energy, and control domains are coupled so that stability, the energy-production rate, reduced inhibition, and reduced failure risk are optimized simultaneously \cite{hamelers2011butler}.

\subsection{Breeding–Induced Mutagenesis}

In breeding based on induced mutagenesis, the dose and type of mutagen, genotype/population, phenotype, growth environment, breeding selection, and performance evaluation operate in a multistage and interdependent manner \cite{oladosu2016principle, holme2019induced, ma2021classical}. In TDDT, mutagenesis is incorporated into the shared state as a change in the genetic composition of the population, and its effects on traits such as yield, resistance, quality, adaptability, and stability are predicted alongside environmental conditions and the breeding objective \cite{oladosu2016principle, holme2019induced, ma2021classical}. This framework can support the selection of promising genotypes, the reduction of costly trials, and the targeted refinement of the selection pathway through genotype–phenotype–environment–decision coupling.

\subsection{GNSS-Online-Independent Navigation in UAV Systems}

In UAV navigation without online dependence on GNSS, position and trajectory estimation depends on the coupling of multiple domains, such as inertia, altitude, heading, wind, airspeed, energy, actuators, flight dynamics, map, obstacles, and safety constraints; an error in any domain can be transferred covertly to other domains and subsequently return to the trajectory estimate\cite{gyagenda2022review, duberg2020ufomap, oleynikova2017voxblox}. In TDDT, these domains are incorporated into a shared state and fast/meso/slow temporal loops so that the relative trajectory, collision risk, energy consumption, wind effects, maneuvering limitations, and map matching are evaluated in a coupled manner. Such high-risk and sensor-limited applications demonstrate that using multiple domains and independent estimation pathways can make the detection of hidden loop flow and the online correction of navigation errors more reliable.

\section{Advantages, Challenges, and Limitations}
\label{sec:advantages}

\subsection{Advantages: Integrated Decision-Making, Trans-Domain Prediction, and Risk Reduction}

In conventional digital twins, the primary objective is often monitoring, simulation, prediction, optimization, control, and decision support for a specific asset, process, or system \cite{jones2020characterising, yao2023systematic, fuller2020digital, botinsanabria2022digital}. However, in TDDT, multiple domain twins, such as climate, biology, energy, economy, infrastructure, health, risk, or control, are placed within a progressive temporal coupling state, and the output of each domain can modify the constraint, input, error, or objective of another domain, thereby enabling integrated decision\-making \cite{budiardjo2021digital, vergara2024federated}.

In complex systems, an error, crisis, or environmental change usually begins in one domain, but its consequences can be transferred to other domains. In general, digital twins can reduce the cost and risk of direct testing on the real system by creating a virtual environment for simulation, monitoring, prediction, optimization, and control, while enabling scenario testing and the prediction of future behavior \cite{yao2023systematic, botinsanabria2022digital}. TDDT extends this capability to the multi-domain level. Consequently, TDDT can be used for risk reduction, preventive decision-making, “what-if” scenario analysis, stress testing, and enhancing the resilience of complex systems \cite{budiardjo2021digital, vergara2024federated, ivanov2023intelligent, brucherseifer2021digital, jones2020characterising,yao2023systematic, fuller2020digital, botinsanabria2022digital}.

\subsection{Challenges of Data, Interoperability, and Standardization}

The most difficult challenge in TDDT is the management of heterogeneous and multi-source data. Digital twins depend on sensor data, event data, models, simulations, historical data, operational data, and sometimes human data for their continuous updating. In TDDT architectures, these data enter the system from different domains and with different temporal rates, units of measurement, accuracy, quality, and semantics. Issues such as data integration, data quality, data discovery, data search, storage, processing, and interoperability remain among the main challenges in DT implementation \cite{correia2023data}. In TDDT, this challenge becomes more severe, as each data item must be interpretable not only within its own domain but also within the shared trans-domain state.

The second challenge is interoperability and standardization. “Data ownership,” “integration among virtual entities,” “levels of fidelity,” “technical implementation,” and “application throughout the lifecycle” are among the major gaps in the DT literature \cite{jones2020characterising}. Domain twins may be developed using different modeling languages, temporal scales, data structures, APIs, ontology, and levels of fidelity. Therefore, TDDT requires adherence to common standards for data definition, identification, semantics, timing, uncertainty, model exchange, V\&V\footnote{Verification and Validation}, and lifecycle management \cite{jones2020characterising, correia2023data, shao2023credibility, budiardjo2021digital, iso2021iso23247, karabulut2023ontologies}. Without such standards, TDDT becomes a collection of separate twins that are connected only superficially but cannot operationally produce a valid and verifiable joint decision \cite{shao2023credibility, budiardjo2021digital}.

\subsection{Challenges of Modeling, Validation, and Uncertainty}

Each domain twin must be based on a physical, mechanistic, data-driven, statistical, agent-based, machine-learning, or hybrid model. In TDDT, the issue is not merely the validation of a single model; rather, the validity of the coupled models, the direction of influence among domains, error propagation, and output consistency must also be examined. The DT literature indicates that the levels of fidelity, credibility, validation, and verification are key issues in digital twins \cite{jones2020characterising, shao2023credibility}. Moreover, in safety-critical twins, particularly in precision medicine, VVUQ is fundamentally important for establishing trust in predictions and decisions \cite{shao2023credibility, sel2025survey}.

Uncertainty in TDDT is multilayered. Part of the uncertainty arises from sensors, incomplete data, delay, noise, and data quality; part originates from model parameters, physical approximations, model structure, training data, and simulation error; and another part emerges from coupling among domains and human decision-making. Therefore, in addition to predicting output values, TDDT must also report the confidence level, uncertainty interval, decision sensitivity, and probability of error propagation among domains. If these uncertainties are not made transparent, TDDT may produce a decision that appears precise but is not scientifically or operationally reliable \cite{shao2023credibility, sel2025survey, deantoni2024quantifying, nist2024digital}.

\subsection{Ethical, Security, Data-Ownership, and Trustworthiness Challenges}

The fifth challenge of TDDT is trustworthiness, security, privacy, and data ownership. Digital twins typically depend on continuous, sensitive, operational, and sometimes personal or confidential data. In TDDT, this issue becomes more complex because data are exchanged among multiple domains, multiple stakeholders, and multiple systems. Jones et al. identify data ownership as one of the key gaps in the realization of DTs \cite{jones2020characterising}. Security and privacy studies also indicate that DTs face challenges such as confidentiality, integrity, authentication in IoT/IIoT\footnote{Industrial Internet of Things} communications, cyberattacks, access control, data leakage, communication security, and trust in the model \cite{elhajj2024systematic, kustelega2024privacy}.

From an ethical perspective, TDDT can generate decisions that affect humans, the environment, the economy, health, or infrastructure; therefore, transparency, explainability, accountability, auditability, human control, and decision fairness must be incorporated into it. In the healthcare domain, ethical studies of DTs indicate that the lack of empirical validation, ambiguity regarding the actual value of DTs, bias, privacy, data ownership, and governance challenges can widen the gap between technological promises and real-world implementation \cite{burr2026realising, huang2022ethical}. This warning is also valid for TDDT because a trans-domain decision may create unintended effects in other domains. Therefore, TDDT must be designed, validated, and monitored not only from technical perspectives but also from ethical, security, legal, and social perspectives \cite{kustelega2024privacy, burr2026realising, huang2022ethical}.

\section{Future Research Agenda for TDDT}
\label{sec:future}

The following agenda comprises a set of proposed directions inferred from the gaps and limitations of this article and does not represent an established scientific consensus. The priority of each direction depends on the application, risk, data, computational resources, and maturity of the domain twins \cite{jones2020characterising, michael2022integration, david2025interoperability, fuller2020digital, khedr2025composition}.

First, the definition, classification boundary, and compliance conditions of TDDT should be examined in independent studies, and the stability of couplings, temporal loops, and error transfer should be tested \cite{heindl2022structured, dalibor2022cross, michael2022integration, david2025interoperability, vergara2024federated, khedr2025composition}. At the same time, a reference architecture, vocabulary, API, provenance mechanisms, and coupling contracts should be developed, and conceptual mappings to standards should be transformed into conformance tests \cite{ david2025interoperability, modelica2024fmi302, ieee2025hla1516, budiardjo2021digital, correia2023data, iso2021iso23247, karabulut2023ontologies}.

Second, open and multi-domain benchmarks should be established to evaluate model validity, the shared state, coupling, decision-making, and end-to-end performance. This evaluation should include baselines, ablation, stress testing, and VVUQ \cite{thelen2022comprehensivepart2, shao2023credibility, sel2025survey, deantoni2024quantifying, nist2024digital}. The effectiveness of single-episode training, CRP, SARG, MRG, KStore, hidden loop current, and online adaptation should also be examined independently; at present, their benefits remain research hypotheses \cite{bolender2021self, kennedy2001bayesian, thelen2022comprehensivepart2, deantoni2024quantifying}.

Third, real-time execution should be investigated through approaches such as FFD, ROM, and lightweight surrogates, without disregarding reductions in accuracy and changes in uncertainty \cite{titsias2009variational,manousakas2020bayesian,polke2026adaptive,semenova2025case,birk2022automatic}. Safety, fallback, human override, security, data ownership, explainability, and decision accountability should also be incorporated from the beginning of the design process \cite{shao2023credibility, elhajj2024systematic, kustelega2024privacy, burr2026realising, huang2022ethical}. Finally, the practical validation of TDDT should proceed progressively from simulation to laboratory environments and then to field pilots; the results of each stage are generalizable only within the conditions under which they were tested \cite{jones2020characterising, shao2023credibility, sel2025survey}.

\section{Conclusion and Answers to the Research Questions}
\label{sec:conclusion}

\textbf{Q1}. This article demonstrated that TDDT extends beyond single-domain, multi-domain, and Cross-Domain DTs because, rather than relying on comparison or reuse across domains, it is based on operational coupling and mutual influence among domains.

\textbf{Q2}. We explained that, within the classification of digital twins, TDDT is not an independent functional level but rather an architectural subtype of a Composite/Federated Digital Twin System-of-Systems for complex and multiscale systems.

\textbf{Q3}. The conceptual architecture of TDDT should connect data, model, state, error, objective, decision, and control through a shared trans-domain state and a coupling layer among heterogeneous domain twins.

\textbf{Q4}. We explained that TDDT, through fast, meso-scale, and slow loops, together with offline learning, online tuning, and feedback from the real environment, can dynamically and adaptively support trans-domain decision-making.

In conclusion, TDDT is an architectural response to the limitations of individual, multi-domain, and CDDT in complex systems. By establishing a trans-domain fused model, a shared state, operational coupling, and the TDOC orchestration core, this approach transfers decision-making from the level of separate domains to the level of the entire system. Single-episode offline training, together with CRP, SARG, MRG, and high-importance datasets, provides the prior knowledge required to reduce the risk and cost of online learning; meanwhile, the offline error function, uncertainty estimation, feedback from the real environment, and the detection of hidden error flows provide the basis for online adaptation. The main functions of TDDT are prediction, scenario generation, risk reduction, decision refinement, and trans-domain control based on real-world feedback. However, its practical realization depends on the standardization of terminology and interfaces, field validation, uncertainty management, data security, computational-cost control, and the development of lightweight and reliable models.

\bibliographystyle{unsrtnat}  
\bibliography{references}

\end{document}